\documentclass{svjour2}[]
\smartqed

\usepackage[english]{babel}
\usepackage{hyperref}
\usepackage{color}
\usepackage[usenames,dvipsnames]{xcolor}
\usepackage{amsmath,amsfonts, slashed, amssymb, wrapfig, subfigure}
\usepackage[latin1,ansinew]{inputenc}
\usepackage{fullpage}
\usepackage[squaren,Gray,thinspace,thickqspace]{SIunits}
\usepackage{graphicx}
\usepackage{verbatim}
\usepackage{morefloats}
\usepackage{blkarray}

\hypersetup{colorlinks=true, citecolor=orange, urlcolor=cyan, linkcolor=blue}
\newcommand{\reff}[1]{(\ref{#1})}
\newcommand{\beq}{\begin{equation}}
\newcommand{\eneq}{\end{equation}}

\newcommand{\fig}[1]{Fig.~\reff{#1}}
\newcommand{\eq}[1]{Eq.~\reff{#1}}

\begin{document}

\title{From Ecology to Finance (and Back?): \\Recent Advancements in the Analysis of Bipartite Networks}
\author{{Mika J. Straka} \and {Guido Caldarelli} \\ {Tiziano Squartini} \and {Fabio Saracco}}
\institute{Mika J. Straka \and Tiziano Squartini \and Fabio Saracco
	\at IMT School for Advanced Studies Piazza San Francesco 19, 55100 Lucca, Italy\\
	\email{mika.straka@imtlucca.it}
	\and Guido Caldarelli 
		\at IMT School for Advanced Studies Piazza San Francesco 19, 55100 Lucca, Italy
		\at Istituto dei Sistemi Complessi, CNR, Dip. Fisica Universit\`a ``Sapienza'', P.le A. Moro 2, 00185  Rome, Italy
		\at London Institute of Mathematical Sciences, 35a South St, Mayfair London W1K 2XF, UK
		\at European Centre for Living Technology, University of Venice, San Marco 2940 30124 Venezia, Italy
	}
\date{\today}

\maketitle

\begin{abstract}	
	Bipartite networks provide an insightful representation of many systems, ranging from mutualistic networks of species interactions to investment networks in finance. The analysis of their topological structures has revealed the ubiquitous presence of properties which seem to characterize many - apparently different - systems. Nestedness, for example, has been observed in plants-pollinator as well as in country-product trade networks. This has raised questions about the significance of these patterns, which are often believed to constitute a genuine signature of self-organization. Here, we review several methods that have been developed for the analysis of such evidence. Due to the interdisciplinary character of complex networks, tools developed in one field, for example ecology, can greatly enrich other areas of research, such as economy and finance, and vice versa. With this in mind, we briefly review several entropy-based bipartite null models that have been recently proposed and discuss their application to several real-world systems. The focus on these models is motivated by the fact that they show three very desirable features: analytical character, general applicability and versatility. In this respect, entropy-based methods have been proven to perform satisfactorily both in providing benchmarks for testing evidence-based null hypotheses and in reconstructing unknown network configurations from partial information. On top of that, entropy-based models have been successfully employed to analyze ecological as well as economic systems, thus representing an ideal, interdisciplinary tool to approach the study of bipartite complex systems. 
As an example, the application of an appropriately defined null model has revealed early-warning signals, both in economic and financial systems, of the 2007-2008 world crisis; another interesting application of the entropy formalism is provided by the detection of statistically-significant export specializations of countries in international trade, a result that seems to reconcile Ricardo's hypothesis in classical economics with the recent findings about the (empirical) diversification of the national productions.
	\keywords{complex networks \and mutualistic networks \and bipartite networks \and ecology \and trade \and financial networks \and systemic risk  \and nestedness \and bipartite motifs \and checkerboards \and null models \and exponential random graphs \and network projection \and network validation \and network filtering}
\end{abstract}

\section{Introduction}
``Data is the New Oil'' has become the unofficial slogan for the enthusiasts of technological progress in the recent decade~\cite{wired2014}. New data sources have created new economic and political possibilities. Catalyzed by the need for new analytical and numerical tools, the theory of \emph{complex networks} has gained much attention, since the interplay between different data agents can often be expressed in the shape of a network, and new methods for the analyses of such structures have been designed. 

A prominent network type found in many real-world systems is the so-called \emph{bipartite network}, which is characterized by the presence of two different types of nodes. Examples are user-movie data bases, plant-pollinator ecosystems, author-article collaborations, or financial bank-asset networks. Although purely data-based analyses provide valuable insight into the mechanisms of networks, recent results have shown that such structures contain more information than is apparent at first sight. In particular, several techniques have been designed based on statistical physics and information theory, which provide the possibility to filter statistically relevant signals from the network that otherwise remain hidden when the data is take at face value~\cite{Saracco2016,Saracco2016a,Straka2017,Gualdi2016,Squartini2017}.

Network theory is by nature interdisciplinary and has created a vast vocabulary and a plethora of tools. Due to the interaction patterns of many biological systems, the analysis of bipartite networks has been very popular in ecology and its methodologies have spread to other ares of research. We present a brief review of insights that have been gained in the areas of ecological networks, economic and financial networks. Our focus, however, lies on bipartite network modelling, with a particular attention to \emph{entropy-based null models} and their applications. We will show that seemingly genuine network characteristics, such as nestedness, can be traced back to basic properties like the degree sequence of the nodes. For this purpose, appropriately defined benchmarks models are used that are as unbiased and general as possible. Furthermore, we review how such null models can be used to reconstruct networks when only partial or noisy data is available, and how this method can be applied to asses systemic risk in financial networks~\cite{diGangi2015,Squartini2017,Gualdi2016}. In addition, we illustrate how Ricardo's specialization hypothesis in international trade~\cite{Ricardo1817} can be reconciled with the apparently contradicting export diversification signal that has been observed~\cite{Strauss-kahn2011}, and how early warning signs are revealed preceding the financial crisis of 2008~\cite{Saracco2016a}. 

Network theory has found wide-spread applications in different fields of scientific research. In this article, we focus on \emph{bipartite networks}, in which we can distinguish between two distinct types of nodes. They be ordered in two separate layers such that links only exists between, but not within layers. The structure of the network can be expressed in a \emph{biadjacency matrix}, with nodes of one type along the rows and nodes of the other type along the columns.
Due to the ubiquity of network structures in science, different fields have created different vocabularies. For instance, in ecology the biadjacency matrix is commonly known as the \emph{interaction matrix} when the interaction between species is studied. Analogously, research on the occurrence of organisms in different environments uses the \emph{presence-absence matrix}. Furthermore, in economic and financial networks one may use the expression \emph{ownership matrix}. We will use the general term biadjacency matrix in the following. Regarding the number of connections attached to each node, we refer to them as \emph{degrees}, which is more common than \emph{marginal totals} in ecology. 

In the following sections, we review major insights that emerged from the study of bipartite networks. We focus initially on ecological networks, which have provided tools and methodologies that have been applied subsequently in other fields, and illustrate results from economic and financial networks. 

\section{Ecological Networks}
\label{sec:ecologicalNetworks}

The analysis of networks has a long tradition in the field of biology and ecology. Research on food webs, for instance, can be dated back to the pioneering works of Elton in 1927~\cite{Elton1927}. Food webs capture the predator-prey relationships between different species: squirrels feed off plants but are hunted by snakes, which fall prey to foxes. Directed links in these networks express the flow of biomass, and species can be order in hierarchical layers (known as trophic levels) according to their position in the food chain.

Ecology focuses on special types of webs and studies the interactions among species, or between species and their natural environments. Some typical examples are plants and pollinators, or organisms and their habitats. In these cases, one can distinguish between two different types of nodes that populate two distinct layer of a bipartite network. If the interactions between the species or environments are mutually beneficial and cooperative, for example in the case of pollinators and plants, such bipartite networks are often referred to as \emph{mutualistic networks}.

\subsection{Bipartite Motifs}
Motifs are defined as $n$-node subgraphs that are overrepresented in empirical networks and have been labeled as ``the building blocks of complex networks''~\cite{Milo2002}. In directed networks, such as food webs, the smallest nontrivial motifs can be built out of three nodes, leading to 13 distinct patterns~\cite{Milo2002}. Different motifs are assumed to serve different functions in the network. In genetic transcription networks, for example, is has been observed that certain motifs regulate the expression of genes~\cite{Alon2007} (for an overview of the motifs and their function, see, e.g., ~\cite{Shoval2010}).

Analyzing networks from ecology, engineering, biochemistry and neurobiology, in~\cite{Milo2002} it has been observed that different network types show distinct motif abundances. Hence, the question arises whether one can predict global network characteristics from the presence and temporal changes of such structures. Finding motifs in monopartite networks can generally be computationally intensive and different algorithms have been proposed (see~\cite{Wong2011} for a survey).

Here, we will concentrate on bi-cliques, i.e. motifs in undirected bipartite networks. We will use the vocabulary presented in~\cite{Saracco2015a}, since, in our opinion, the nomenclature makes it easier to grasp the shape of the motifs. As an example, the M-, W-, X-, as well as the V-, $\Lambda$-, and the V$_n$-motifs, are shown in \fig{fig:motifs}.
\begin{figure}[t]
	\begin{center}
		\includegraphics[width=0.5\textwidth]{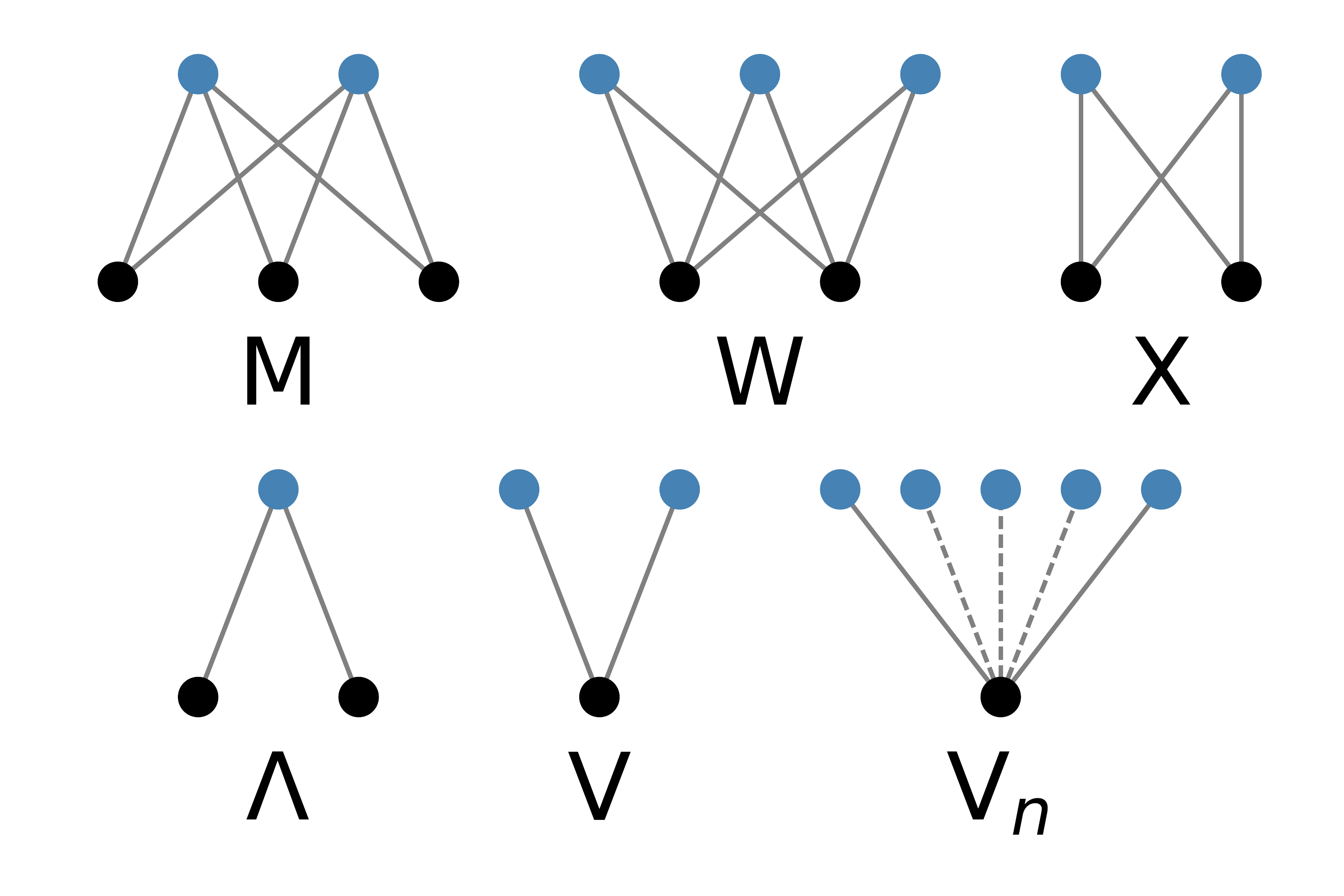}
	\end{center}
	\caption{Illustration of some undirected bipartite motifs. The nomenclature is based on the visual shape of the structures. \textbf{Top:} closed motif in which all nodes of one layer are connected to those of the other. \textbf{Bottom:} open motif that capture node similarities in terms of common nearest neighbors in the opposite bipartite layer.}
	\label{fig:motifs}
\end{figure}

The simplest motifs are the bi-cliques $K_{1,2}$ and $K_{2,1}$, also known as
$\Lambda$- and V-motifs, that are composed of two nodes in the same and one
node in the opposite layer. They draw exactly a ``$\Lambda$'' and a ``V'' between the layers, as illustrated in Fig.~\reff{fig:motifs}. 

Since the network is describe by a binary biadjacency matrix, we can easily express the number of V-motifs between the nodes $i$ and $j$ of the upper layer as
\begin{equation}\label{eq:V}
V^{ij}=\sum_{\alpha\in\text{L}} m_{i\alpha}m_{j\alpha},\quad i, j\in \Gamma.
\end{equation}  
The $\Lambda_{\alpha\beta}$-motifs are defined analogously for the nodes of the lower layer, $\alpha, \beta\in$ L. V$^{ij}$ captures the number of
neighbors that the node couple $(i, j)$ has in common. The motifs can be easily
generalized to more than two nodes by including $n$ legs that are all attached
to the same node in the opposite layer, as shown in \fig{fig:motifs}. We will call them $V_n$ and $\Lambda_n$
(with $\text{V}=\text{V}_2$ and $\Lambda=\Lambda_2$), or in
standard graph theory $K_{2,n}$ and $K_{2,n}$. V- and $\Lambda$-motifs
thus represent the number of connections shared between 2 or more nodes belonging to
the same layer.  

A more complex class of motifs is represented by the
so called \emph{closed motifs}. The M-, W- and X-motifs are illustrated on the top in Fig.~\reff{fig:motifs} and are referred to as $K_{2,2}$, $K_{3,2}$ and $K_{2,3}$ in graph theory. We can express them in terms of the biadjacency matrix and write, for instance, for the total number of X-motifs
\begin{equation}\label{eq:X}
X=\sum_{i<j}\sum_{\alpha<\beta}m_{i\alpha}m_{j\alpha}m_{i\beta}m_{j\beta}.
\end{equation}
The other mentioned closed motifs can be described similarly. 

Bipartite motifs can even account for non-existing links, which is the case, for example, of the popular \emph{checkerboards}, introduced by Diamond  \cite{Diamond1975} for the study of the avifauna of the Vanuatu islands. A checkerboard considers the case of mutual exclusions of two species. The total number of checkerboards is in the biadjacency matrix is thus
\begin{equation}\label{eq:C}
C=\sum_{i<j}\sum_{\alpha<\beta}m_{i\alpha}(1-m_{j\alpha})(1-m_{i\beta}) m_{j\beta}.
\end{equation}
\emph{Togetherness}, $T$, is defined in a similar way and counts how many times two species interact together with the same species, avoiding, at the same time, the interaction with other ones. In formulas,
\begin{equation}\label{eq:T}
T=\sum_{i<j}\sum_{\alpha<\beta}m_{i\alpha}m_{j\alpha}(1-m_{i\beta}) (1-m_{j\beta}).
\end{equation}
In \cite{Stone1990}, the authors show that $C$ and $T$ differ by a constant term. 

As a final comment to the present section, note that although all the motifs so far involve several links, they are all multi-linear in the corresponding biadjacency matrix. This fact is particular convenient for analytical calculations, as we will see in the following.

\subsubsection{Motifs Analysis in Mutualistic Networks}
In ecology, finding patterns that explain the distribution of species in different habitats, e.g. islands, forest, or even parasitic hosts, or why certain organisms interact with others, is a major concern. In this context, the frequency of motifs permits to highlight hidden structures in the architecture of the biological system.
A clear example is the case of the study of the avifauna of the Vanuatu islands, \cite{Diamond1975}. Considering birds and their island habitats as the two layers of a bipartite network, the abundance of checkerboards is analyzed in order to understand co-existing behaviors. This question triggered a long debate about the correct null model to test statistically significance of the measurements~\cite{Connor1979,Gilpin1982,Diamond1982,Roberts1990}. An agreement was achieved with the work of~\cite{Roberts1990}: others proposed a rewiring randomization in which the degree sequence is fixed for both layers, as well as the average bird population of the islands bird species occupy. The authors observe a statistically significant abundance of checkerboards, suggesting a peculiar colonization pattern that increases the mutual exclusions of some species.

Checkerboards and togetherness in a bipartite network can be measured using, for example, the package released by Dormann et al.~\cite{Dormann2009}.

\subsection{Nestedness}

\begin{figure}[t]
	\begin{center}
		\includegraphics[width=0.5\textwidth]{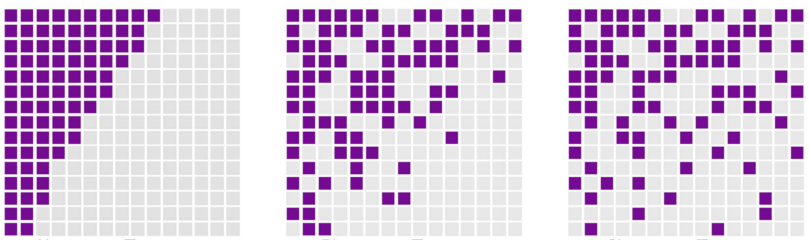}
	\end{center}
	\caption{Illustration of three different matrices of the same dimensions and number of links (filled squares). The left-most matrix can be packed more densely into a triangular shape than the other two and has the highest nestedness. Notice how ``shorter'' rows (columns) are completely contained in ``longer'' rows (columns.)  The nestedness clearly decreases to the right. Image taken from \cite{Johnson2013}.}
	\label{fig:nestedness}
\end{figure}

From the study of ecological systems, the insight has emerged that species in sites of lower biodiversity also populate environments with larger biodiversity. This concept is called \emph{nestedness} 
and translates into the fact that specialists' interactions, i.e. organisms that interact only with a small number of other species, are a subset of those of generalist organisms. 
This property is reflected in the structure of the biadjacency matrix: rows and columns can be sorted in such a way that the matrix is approximately triangular, as shown in \fig{fig:nestedness}. The role of such a structure is debated, as we will see in the following, but nevertheless it is constantly present in different mutualistic or antagonistic system.

During the years, several metrics to capture the nestedness phenomenon have been proposed in literature, with the first attempt dating back to the \emph{nestedness temperature}~\cite{Atmar1993}. After ordering rows and columns in the biadjacency matrix into a state of ``maximum packing", a line is drawn on the matrix representing the boundary of the expected fully nested matrix. Then, a quantity called ``temperature'' is defined by considering the absence in the packed part and the presence in the empty side of interactions, weighted by their distance from the boundary.  In \cite{Almeida-Neto2008}, the authors show that the nestedness temperature is not maximal for disordered system, since random matrices have a intermediate value of nestedness, and proposed the NODF (``Nestedness metric based on Overlap and Decreasing Fill'') to solve the problem. 
NODF is independent of the order of the elements in the matrix and is defined as a weighted sum of row and column contributions. More in details, this quantity gets a contribution from every couple of elements belonging to the same layer and, if they have different degrees, it counts the number of common interactions. Some scholars argued about the opportunity of disregarding the contribution of couple of nodes with the same degree and, for instance, Bastolla et al. \cite{Bastolla2009} provided a different measure considering this contribution.

The role of nestedness for the properties of ecological networks has been debated. On the one hand, it has been argued that nestedness generally increases biodiversity by reducing competition~\cite{Bastolla2009} and favors the stability of the network~\cite{Thebault2010}. On the other hand,~\cite{Staniczenko2013} claim that nested interactions are inherently less stable compared to random interaction. Although predator-prey interactions seem to stabilize the networks, mutualistic and competitive interaction do not~\cite{Allesina2012}. Note that the presence of loops ensure redundancy in ecological networks~\cite{garlaschelli2003universal,palamara2011population} but might trigger instability in financial networks~\cite{Bardoscia2017}
An important contribution to the discussion was put forward in~\cite{Suweis2013}: the authors show that the attempts of a species to increase its abundance in a mutualistic network drives the system to a more nested configuration. In this scenario, species abundances start from general initial conditions and growth is shown to be higher if the number of mutualistic interactions is lower.
Moreover, the abundance of the rarest species is connected to the resilience of the network, i.e. the speed at which the system, after small perturbations, returns to an equilibrium.

Despite the efforts, no consensus about the importance of nestedness has yet been reached. James et al.~\cite{James2012} show that the correlation between persistence and nestedness is present when nestedness correlates even with the connectance of the network. Hence, it is not clear which variable should be considered among connectance and nestedness. 
A possible reason for this observation has been presented by Johnson et al.~\cite{Johnson2013}, who argue that nestedness naturally derives from degree heterogeneities and disassortative degree-degree correlation, i.e. the tendency of high degree nodes to connect to low degree nodes. As they point out, finite null models, such as the widely used \emph{Configuration Model} (CM, ~\cite{Molloy1995,Newman2001,Chung2002,park2004statistical}) tend to be dissassortative and nested. They conclude that in almost 90\% of their 60 studied real empirical networks, the nestedness can be described by a degree-conserving null models. 

As highlighted by Johnson et al.~\cite{Johnson2013}, a null model should be implemented in order to state if the nestedness is a genuine quantity or it is already captured by the degree sequence. The authors choose the configuration model in the version of ~\cite{Chung2002}, which is valid for sparse networks (as most of the mutualistic networks), but performs poorly on more dense systems. In the following subsection, we will introduce a more general class of null model and its simplest realization, the Bipartite Random Graph, and show and how it was employed to uncover non trivial properties of mutualistic networks. In the Economic Networks section we will see how such a framework can be generalized to embed the information of the degree sequence, while the Financial Networks section will show how it can be generalized to capture the information of a weighted network.

\subsection[]{Bipartite Exponential Random Graph I. -- Bipartite Erd{\H o}s-R{\'e}nyi Random Graph}
Statistical null models can be used as comparison benchmarks in order to verify whether real systems show unusual properties. For this purpose, they should be unbiased and formulated as general as possible. This notwithstanding, null models may maintain certain characteristics of the empirical network that should be discounted in the construction of the ensemble. In the area of complex networks, one of the principle aims is to obtain probability distributions for different graph instances, preferably factorisable in terms of link probabilities for analytical tractability.

In these sections, we review the extension of the exponential random graph to bipartite networks,
motivated by the fact that it monopartite version has enjoyed considerable success in the past~\cite{Jaynes1957,park2004statistical,Garlaschelli2008}. We will refer to the bipartite framework as the \emph{bipartite exponential random graph model} (BERG).  More detailed derivations of the null models can be found in the Appendix~\reff{sec:appendixNullModels} and in~\cite{Saracco2015a,Saracco2016,Straka2017}.

Let us consider an empirical binary bipartite network, expressed by its biadjacency matrix $\bold{M}^*$ with layer dimensions $N_\text{L}$ and $N_\Gamma$. Quantities measured on the real system will be marked with an asterisk.
We start by constructing the set of all possible networks with the same layer dimensions: this set, the \emph{ensemble} $\mathcal{G}_B$, runs from the empty network (without any links) to the fully connected network (in which all possible $N_\text{L} \times N_\Gamma$ links are realized). 
To every member of the ensemble we will assign a probability that depends on some property of the original network that we want to maintain. By comparing the ensemble characteristics with the original network, we can observe how certain information constraints on the ensemble capture features of the real system: if the quantities measured on the real network are correctly reproduced by the ensemble, the imposed constraints are enough to explain such a behavior. If, however, the real network shows a statistically significant deviation, this information cannot be explained by the constraints and represent a non-trivial information about the structure of the empirical network.

Be $G_B \in \mathcal{G}_\text{B}$ an element of the ensemble $\mathcal{G}_\text{B}$ of bipartite
networks with fixed layer dimensions. The most general and unbiased probability distribution over the
ensemble can be obtained by maximizing the Shannon entropy~\cite{park2004statistical},
defined as 
\begin{equation}\label{eq:S}
S=-\sum_{G_\text{B}\in\mathcal{G}_\text{B}} P(G_\text{B}) \ln\big(P(G_\text{B})\big).
\end{equation}
Assume now that we have measured some quantities $\vec{C}$ of the real network, for example the number of edges. We want the corresponding ensemble expectation value $\langle\vec{C}\rangle$ to reflect the same value, i.e. we constrain the expectation value of the observable in such a way that $\langle\vec{C}\rangle \equiv \vec{C}^*$. It can be shown that the probability of observing the generic graph $G_\text{B}\in\mathcal{G}_\text{B}$ is, as in \cite{Jaynes1957},
\begin{equation}
P(G_\text{B}|\vec{\theta})=\dfrac{e^{-\vec{\theta}\cdot\vec{C}(G_\text{B})}}{\mathcal{Z}(\vec{\theta})},
\end{equation}
where $\vec{\theta}$ is the vector of Lagrange multipliers associated to the
constraints $\vec{C}$, and $\vec{C}(G_\text{B})$ is the value of the constraints on
the graph $G_\text{B}$.  $\mathcal{Z}(\vec{\theta})$ is the partition function
known from statistical physics,
\begin{equation}
\mathcal{Z}(\vec{\theta})=\sum_{G'_\text{B}\in\mathcal{G}_\text{B}}e^{-\vec{\theta}\cdot\vec{C}(G'_\text{B})},
\end{equation}
and its exponent the graph Hamiltonian, $\mathcal{H} =\vec{\theta}\cdot\vec{C}$~\cite{park2004statistical}. Assuming that the network quantities $\vec{C}$ can be expressed analytically in terms of the biadjacency matrix,  $\vec{C}\equiv \vec{C}(\bold{M})$, we can see that the probability $P(G_B|\vec{\theta})$ for a given graph only depends on the Lagrange multipliers. The trick is to derive their values by maximizing the likelihood of observing the real network in the ensemble, $\mathcal{L}\equiv\ln P(G^*_B)$~\cite{Garlaschelli2008}. This is equivalent to explicitly imposing $\langle \vec{C} \rangle=\vec{C}^*$ on the ensemble.  

The formalism above defines the \emph{bipartite exponential random graph model}, which extends the \emph{exponential random graph model} (ERGM)~\cite{park2004statistical} to networks with bipartite structure. It is well known that constraining the number of links $E^*$ in the ERG framework returns the Erd{\H o}s-R{\'e}nyi random graph~\cite{erdos1959random}. In analogy, imposing the same constraints on the ensemble $\mathcal{G}_B$ gives us the \emph{bipartite random graph} (BiRG, as it is called in ~\cite{Saracco2016}), in which all links have the same probability $p = E / N_L N_\Gamma$. The derivation of the BiRG is shown in the Appendix~\reff{sec:appendixNullModels}.

\subsubsection{Degree Sequence in Bipartite Biological Networks}
Entropy based approaches for the analysis of biological systems percolated almost diffusely in the literature~\cite{Harte2011,Azaele2016a}, but they were rarely employed for the analysis of bipartite networks. Williams~\cite{Williams2011} used the aforementioned BiRG 
to state the significance of the degree distribution in mutualistic networks. The author sampled the ensemble of the BiRG and compared the observed degree distribution with the frequencies expected from the null model by implementing the likelihood ratio statistics. The calculation is repeated for every element of a sample of the BiRG ensemble and the values are compared. The comparison shows that the degree distribution of mutualistic network, besides being strongly skewed, can be usually explained just by the total number of links. The result is even more striking, considering that its monopartite analogous has not such a good performance~\cite{Williams2010}. \\

Summarizing, at the moment we have three observations that cast doubts on the role of the nestedness. On the one hand, the degree distribution is in agreement with the expectation of a null model constraining the total number of links. On the other hand, nestedness is related to the links abundances when it has a positive correlation with the resilience of the network (see~\cite{James2012}). Finally, the degree sequence generally explains the value of the nestedness in mutualistic networks~\cite{Johnson2013}). These three key results give causes to doubt the real role of nestedness in biological mutualistic networks and motivate the need for further analysis.

\subsection{Monopartite Projections and Communities}
\label{sec:stylizedComdet}

When studying mutualistic networks, the question naturally arises whether one can find groups of highly cooperative species, or groups of organisms that compete for the same resources. In plant-pollinator networks, an example for the first case would be a community of plants and pollinators that live in symbiosis and benefit from cooperation. Contrary to that, an example for the latter would be a collection of insects that compete for the same pollen. In ecology, these substructures are referred to as \emph{compartments}~\cite{Thebault2012}. In the following, we shall adopt the network vocabulary and call them \emph{modules} or \emph{communities}. They describe collections of nodes that are more closely related to each other than to individuals in other communities. 

The problem of finding communities between nodes of the same layer can be found throughout different fields of complex networks analysis, from ecological,
to financial, to economic networks. For this problem, several tools have been presented literature (for an overview, see,  e.g.~\cite{Fortunato2010}).   
A popular approach is to perform a \emph{monopartite projection}, i.e. to project the bipartite network on one of its layers. In the resulting graph, nodes are connected if they share at least one neighbor in the original bipartite network.
Note that the procedure discards information -- in general, it is not possible to reconstruct the original bipartite network from the projection. Moreover, there is no clear guideline on how to set link weights in the projection. It has been shown that the communities found in binary projections can be incorrect and misleading and that weighted projections should generally be preferred~\cite{Guimera2008}. Nonetheless, simply setting the weights equal to the number of neighbors in the original network is quantitatively biased~\cite{Zhou2007}. Inspired by the importance of collaborations in the author-article  network of scientific coauthors, Newman proposed that links in the author projection should be corrected by a factor $1 / (d - 1)$, where $d$ is the degree of the collaboration paper~\cite{Newman2001,Newman2004}. Despite these efforts, a systematic exploration of how weight should be set remains open. At the same time, the question of which links carry statistically relevant information is neglected.

\section{Economic Networks}
Seminal works in classical economics date back to Adam Smith's fundamental ``The Wealth of Nations'' in 1776~\cite{Smith1776}. In the wake of Smith's publication, David Ricardo devoted parts of his intellectual endeavors to economics, which culminated in his famous ``Principles of Political Economy and Taxation''~\cite{Ricardo1817}. His most important legacy is probably the concept of \emph{comparative advantage}, which expresses the fact that some nations can produce certain products more efficiently than others. As a result, Ricardo advocated the idea that nations should concentrate their resources only on their most advantageous industries. According to him, combining industrial specialization with free trade would be favorable for all countries and foster national economic growth.

Nowadays, international exportations and importations are recorded on yearly base and made available by the UN Comtrade Database\footnote{Comtrade found at https://comtrade.un.org/}. This allows us to scrutinize trade relations and test hypotheses of classical economics with the help of state-of-the-art tools in data analysis and network theory. In fact, the global structure of trade interactions can be expressed as the so-called International Trade Network (ITN), also known as World Trade Web, in which nodes correspond to countries and link weights to trade volumes in USD. Countries can share directed links with different weights, corresponding to products of different categories.

Trade is one of the main global stages on which countries interact, and the ITN has been extensively studied due to its importance for economic growth and to address questions like globalization and the spreading of economic shocks~\cite{Duenas2013}. For example, regarding the number of trade partners, it has been shown that the network is generally disassortative, i.e. that countries with many trade partners tend to interact with nations with only few ones~\cite{Serrano2003,Garlaschelli2004}. When trade volumes are taken into account, however, it has been observed that high-degree countries trade most intensively with other high-degree countries~\cite{Fagiolo2009}.
Although product-specific trade volumes are very heterogeneous~\cite{barigozzi2010multinetwork}, the aggregate link weights distribution is almost log-normal~\cite{barigozzi2010multinetwork,Annunziata2015}. Country-specific trade volumes depend strongly on national GDP and their distributions reach from truncated log-normality to Pareto-log-normality~\cite{Annunziata2015}. 

International trade can also be studied at a even finer level, when links are drawn among regional industries instead of countries. Using the World Input-Output Database, it has been shown that global production systems are still regionally organized and industries are asymmetrically connected, leading to possible shock amplification from regional fluctuations to the global scale~\cite{Cerina2014}.

\subsection{Diversification in Trade}
\begin{figure}[t]
	\begin{center}		
		\includegraphics[width=0.4\textwidth]{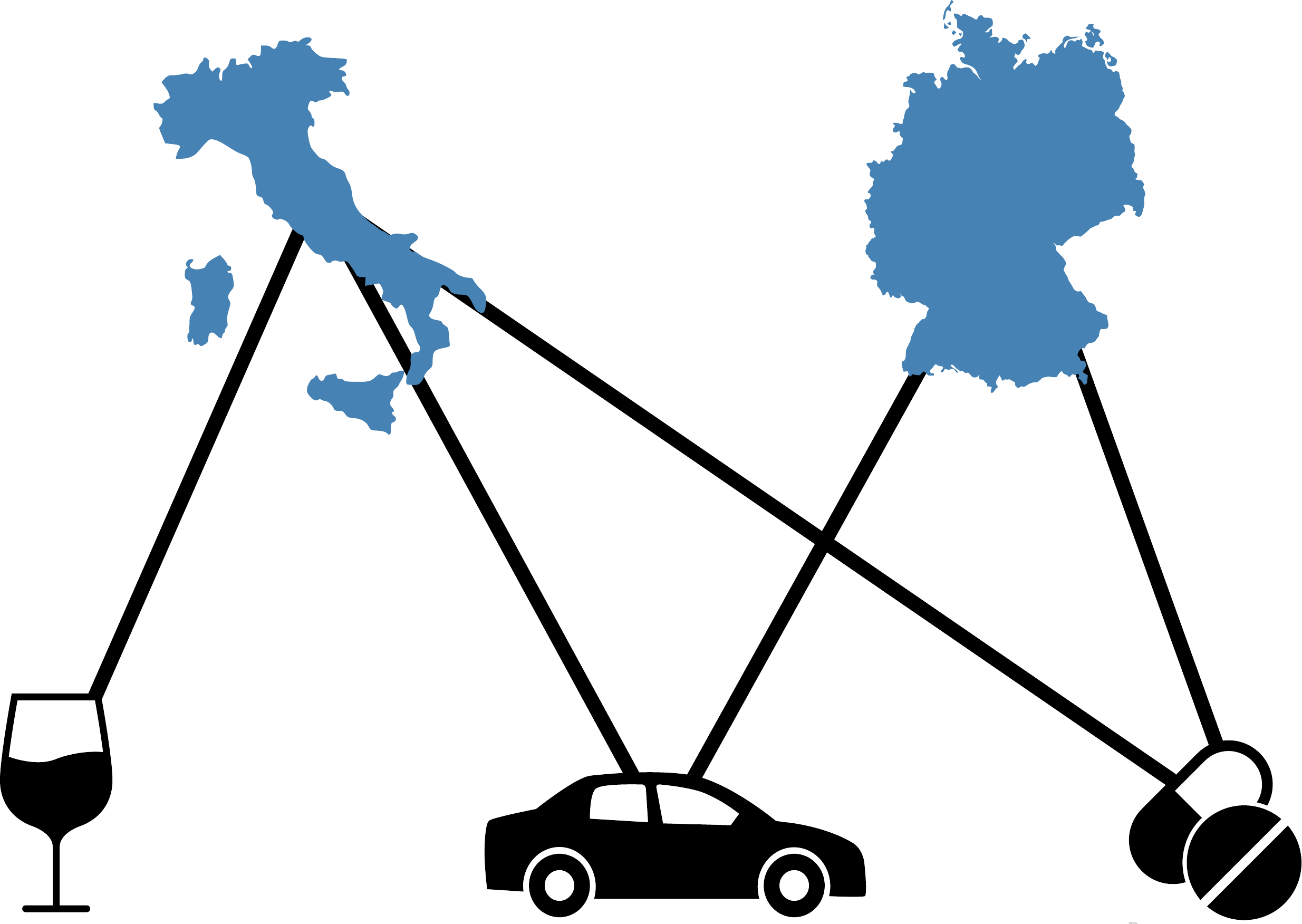}
 	\end{center}
 	\caption{Illustration of a part of the country-product exportation network. Both Italy and Germany are strong exporters of cars and pharmaceutical products, whereas only Italy has a comparative advantage in wines (``Wine'' by Thengakola, ``Car'' and ``pills'' by alrigel from the Noun Project. All icons are under the CC license).}
 	\label{fig:tradeNetwork}
\end{figure}

Recent developments on the ITN have been triggered by the suggestion that trade networks should be considered as bipartite with countries in one layer and products in the other layer~\cite{Galeano2012}. The setup is illustrated in \fig{fig:tradeNetwork}. The proposal is motivated by the observation that importers and exporters have intrinsically different motivations for connecting to trade partners~\cite{Galeano2012}. For the analyses, the authors of~\cite{Galeano2012} made use of methodologies developed for mutualistic networks and analyzed the properties of the country-product network using the \emph{revealed comparative advantage} (RCA), also knows as Balassa index~\cite{balassa1965trade}. The RCA compares the relative monetary importance of a particular product among all exports of a country (its \emph{export basket}) to the global average and assigns a value to each link accordingly, as explained in detail in Appendix~\reff{sec:appendixRCA}. 
 By pruning links sequentially for different RCA threshold values, in~\cite{Galeano2012} the authors separate the core and periphery of the network and show that degree distributions are truncated power laws. 
The networks emerging from the pruning procedure are generally considered as binary, since each existing link expresses the fact that a certain country is a relevant exporter of a particular product at some threshold value. 

\begin{figure}[t]
	\begin{center}
		\includegraphics[width=0.6\textwidth]{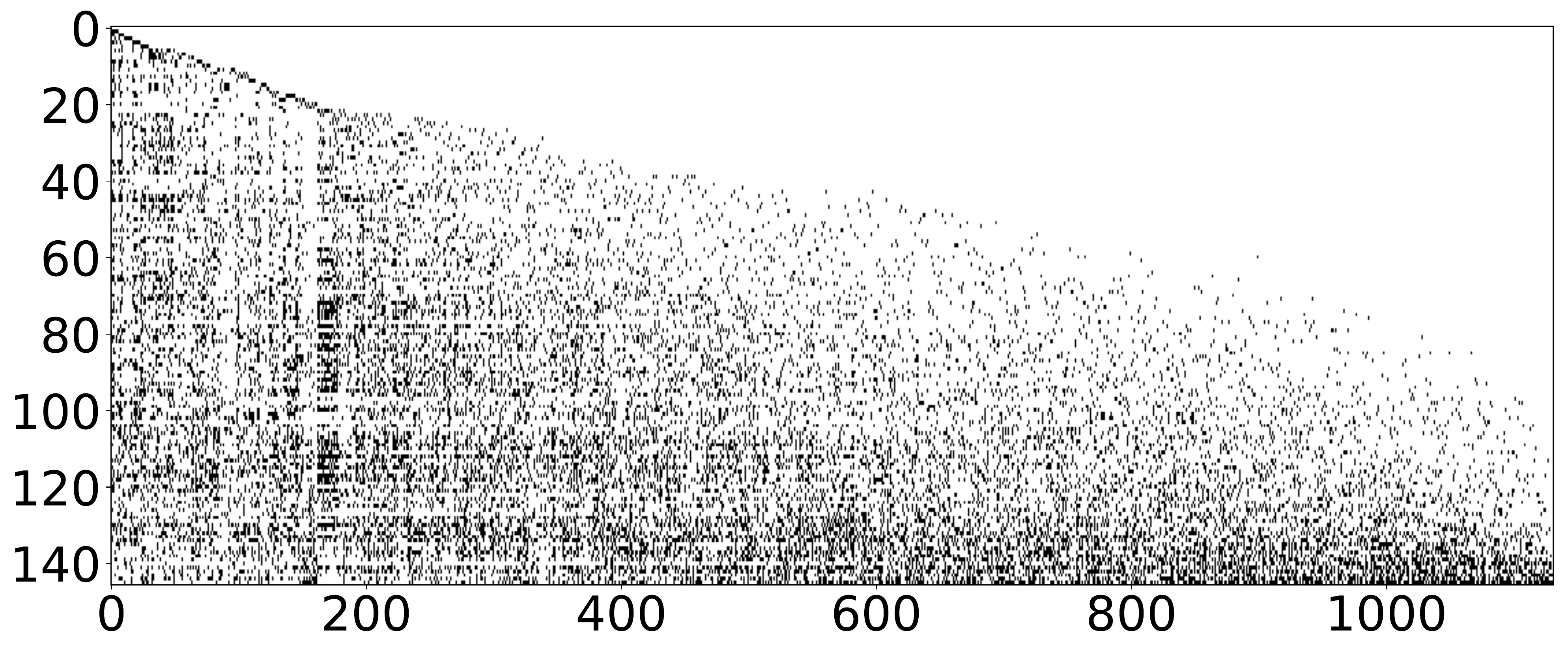}
	\end{center}
	\caption{Biadjacency matrix of the international trade network for the year 2000 with countries sorted from top to bottom and products from left to right in increasing fitness and complexity, respectively. Links in the network are shown as black dots. The overall triangularity of the matrix is correlated with the nestedness of the system.}
	\label{fig:triangularITN}
\end{figure}

A fundamental observation that emerges from the binarized ITN, when only relevant exportation with RCA $\ge 1$ are kept, is the approximately triangular structure of its biadjacency matrix, as illustrated in~\fig{fig:triangularITN}: some countries have large export basket and other small ones, just like some product have only few exporters and others many. The crucial fact is that the smaller export baskets are contained in the bigger ones. The ITN therefore exhibits the nestedness property~\cite{Tacchella2012,Hidalgo2009,Hausmann2011,Cristelli2013,Hidalgo2007,Caldarelli2012,Zaccaria2014},  which we have already observed for mutualistic networks in the previous sections.
In the context of the bipartite trade network, this observation is striking: it contradicts classical economic theories. As mentioned above, according to Ricardo we would expect a specialization of exportation, which should be observable through a block-diagonal structure in the biadjacency matrix. Instead, the matrix is approximately triangular which corresponds to an increasing diversification of exportations, as has also been mentioned in~\cite{Strauss-kahn2011}. The most developed countries export all products, from the most sophisticated to the most basic ones, whereas less developed countries are able to export just few low technology items.

\subsection{Product and Country Space}

A considerable amount of work on the bipartite trade network has been devoted to the analysis of relations among products and among countries. An intuitive approach would be to project the bipartite network on its two layers, respectively. However, this approach is generally problematic -- in fact, in the case of the ITN the projected networks are almost completely connected with link densities of over 93\%~\cite{Saracco2016}, leading to trivial properties.

To address this question, in~\cite{Caldarelli2012} the authors have applied Minimal Spanning Forests to the country and the product projection. Unexpectedly, they find that neighboring countries compete over the same market rather than diversifying their export baskets~\cite{Caldarelli2012}.

A different approach has been chosen by Hidalgo et al.~\cite{Hidalgo2007}, who construct the ``product space'' by connecting products that are similar according to a specific metric. The distance between two products is essentially measured as the conditional probability that a country exports both of them as measured on the data~\cite{Hidalgo2007}. 
They observe that more sophisticated goods, such as vehicles and machinery, occupy the core of the network, whereas less sophisticated ones, e.g. vegetables or crude oil, populate the periphery. Given the topology of the product space, they argue that less developed countries get trapped in the periphery because of a lack of connections to the more prestigious products in the core~\cite{Hidalgo2007}. 

Another proposal for inferring relations among products and for a possible evolution of the industrialization of countries is proposed by \cite{Zaccaria2014}: from the binary bipartite network of trade they are able to obtain a forest of products, discounting the degree sequence of both products and countries. 

All methods revised here do not rely on an unbiased null model, but use different ingredients in order to highlight a possible dynamic for the industrialization of countries. None of them discusses the statistical significance of their findings, but they use some of the features of the bipartite network to propose an explanation for their observations. In order to correctly project the information contained in the bipartite network more involved methodologies are needed.

\subsection{Economic Complexity}
The bipartite structure of the ITN should encode information about non-tradable capabilities of countries~\cite{Angelini2017}, such as their infrastructure, education system, patent rights, and industry-specific knowledge. The fundamental idea is the following: the fact that a country is capable of exporting a certain product signals that its industry is advanced enough to compete in global markets~\cite{Angelini2017}. Consequently, the country has the necessary latent capabilities to manufacture the product. 

In order to capture the complexity of a national economies, Hidalgo and Hausmann proposed the so-called \emph{method of reflections}~\cite{Hidalgo2009,Hausmann2011}. Essentially, the method consists of iteratively assigning a quantity to each node that depends on those of its neighbors and their degrees. As the authors point out, the resulting ``complexities'' of countries correlate with their GDPs. Unfortunately, the convergence of the algorithm is not be guaranteed~\cite{Pugliese2016}.

This problem was remediated in~\cite{Caldarelli2012,Tacchella2012,Cristelli2013} and a non-linear recursive algorithm was proposed, which gave rise to the so-called \emph{economic complexity} framework. The capabilities of countries were labeled as their \emph{fitness} and the level of sophistication of the products as their \emph{complexity}. Although some convergence issues were still present, it has been shown that fitness and complexity rankings of countries and products are stable even in absence of convergence~\cite{Pugliese2016}. 

As already observed for the method of reflections, national fitness seems to correlate with national GDP~\cite{Hidalgo2009}. Accordingly,~ \cite{Cristelli2015} studied the evolution of countries in terms of their fitness (intangible assets assessing competitiveness) and GDP per capita (GDPpc, a monetary measure). They observe a strong heterogeneity in the country dynamics and identify several regimes, such as a ``poverty trap'' in the low fitness regime, and a laminar region for high fitness countries. In conclusion, they argue that the overall heterogeneous evolution dynamics cannot be assessed with classical regression tools and that methods from dynamical systems theory would be more appropriate~\cite{Cristelli2015}. 

In a recent study, the evolution of products has been analyzed in an analogous way~\cite{Angelini2017}. Similar to countries, the dynamic of products is observed in the complexity-logPRODY space, with logPRODY being a monetary measure defined as the average weight of a product exporter's GDPpc~\cite{Angelini2017}. As the authors observe, products tend to move towards an asymptotic zone with product-specific asymptotic markets. Interestingly, the asymptotic markets seem to be determined by the product complexities and are characterized by high competition~\cite{Angelini2017}.

At first sight, fitness and complexity seem to be genuine quantities that complement the topological information on nodes. As has been shown~\cite{Saracco2015a}, and will be discussed further on, this may generally not the case: by applying an appropriately defined null model, the fitness and complexity values can be reproduced based solely on the node degrees. 
\\

Even though the study of the international trade network has enjoyed much attention in the last decade, it is striking that no early warning signals indicating the advent of the financial crisis in 2007--2008 have been observed. This comes as a surprise, since financial and trade relation are strongly connected: in the aftermath of the crisis, world merchandise exports fell by 22\%~\cite{WTOWorldTradeOrganization2015}. The absence of such an observation may be due to the commonly applied RCA binarization procedure. If all export baskets are affected in a similar way by the crisis, no salient signal will be detected. We will see in the next section that more subtle effect can be uncovered through the application of a null model.

\subsection[]{Bipartite Exponential Random Graph II. -- Bipartite Configuration Model}

In the previous section about Bipartite Exponential Random Graph we have focused on the simplest constraint possible, i.e. the total number of edges $E$. We proceed by considering the degree sequence in this section. 
In analogously to the attempts of Johnson and Williams to understand the role of nestedness and its relation with the connectance and the degree sequence, here the rationale is to understand how much the information of fitness and complexity is already contained in the degree sequence, if the signal of diversification is genuine and whether it hides some other important features of the entire system. 

In the realm of monopartite networks, the \emph{Configuration Model} (CM, ~\cite{Molloy1995,Newman2001,Chung2002,park2004statistical})
has enjoyed a variety of application. It is constructed using the degree sequence of the real network and can be extended to the bipartite case, giving rise to the \emph{Bipartite Configuration Model} (BiCM,~\cite{Saracco2015a}). Constraining the degree sequence of both layers, L and $\Gamma$, corresponds to imposing two series of Lagrange multipliers, $\vec{\theta}$ and $\vec{\rho}$, respectively. With some algebra, it can be shown that the probability distribution becomes~\cite{Saracco2015a}
\begin{equation}\label{eq:BiCMProbabilityDistribution}
P(G_B|\vec{\theta}, \vec{\rho}) = \prod_{i, \alpha} (p_\text{BiCM})_{i\alpha}^{m_{i\alpha}}\big(1-(p_\text{BiCM})_{i\alpha}\big)^{1-m_{i\alpha}},
\end{equation}
where the probability per link reads
\begin{equation}
(p_\text{BiCM})_{i\alpha}=\frac{e^{-(\theta_i + \rho_\alpha)}}{1+e^{-(\theta_i+\rho_\alpha)}}, \quad i\in\text{L}, \alpha\in\Gamma
\label{eq:bicmProbability}
\end{equation}
The Lagrange multiplicators can be recovered through the maximization of observing the loglikelihood $\mathcal{L}$, as shown in Appendix~\reff{sec:appendixNullModels}.

Note that \eq{eq:BiCMProbabilityDistribution} factorizes into the single link probabilitiess. This is very convenient for analytical calculations, for example for the multi-linear bipartite motifs, such as V in \eq{eq:V} and X in \eq{eq:X}.
\\

Other less strict null models can be defined through the relaxation of the constraints. For instance, imposing only the degree sequence of only one layer leads to the \emph{bipartite partial configuration model} (BiPCM,~\cite{Saracco2016}). The choice of the null model generally depends on the information that one wishes to discount.

\subsubsection[]{Validated Projections}
The BiCM can be used to safely project the information contained in the bipartite network, discounting the information from the degree sequence~\cite{Gualdi2016,Saracco2016,Straka2017}. The idea is to compare the observed co-occurrence of links between nodes on the same layer (or, otherwise stated, the number of V-motifs between nodes on the same layer) with the expectation from the null model. In the BiCM, given a node couple, the probabilities for the V-motifs insisting on them are independent and, in general, different \cite{Saracco2016}. Thus, the distribution of such V-motifs is a Poisson-Binomial, the generalization of a binomial distribution to independent events with different probability \cite{Hong2013}. The comparison between the observations with the expectation of the null model can be captured by a p-value, such that we have one p-value for every node couple: p-values are analyzed through a multiple hypothesis test and only statistically significant V-motif abundances are validated, leading to a monopartite network with only statistically relevant links. This methods is known in literature as the \emph{grand canonical projection algorithm} and is described in detail in~\cite{Gualdi2016,Saracco2016,Straka2017}. 

\subsection{Network Validation in Trade}
In the following paragraphs, we review some recent results that have been obtained through the application of the binary BiCM in the area of international trade. As we will see, discounting the degree sequence of the empirical network has revealed interesting new results. 

To compare the empirical network with the null model, we will make us of z-scores. Be $Q(G_B)$ a quantities that we can measure on the network $G_B$. \eq{eq:BiCMProbabilityDistribution} gives us the tools to calculate its expectation value $\langle Q \rangle$ and the standard deviation $\sigma_Q$ on the BiCM ensemble. The z-score is defined as
\begin{equation}
z(Q) = \frac{Q^* - \langle Q \rangle}{\sigma_Q}
\end{equation}
and expresses the discrepancy between the observed and the expectation value in terms of standard deviations.
\\

\subsubsection{Nestedness and Specialization}
\begin{figure*}[t]
	\begin{center}
		\includegraphics[width=0.6\textwidth]{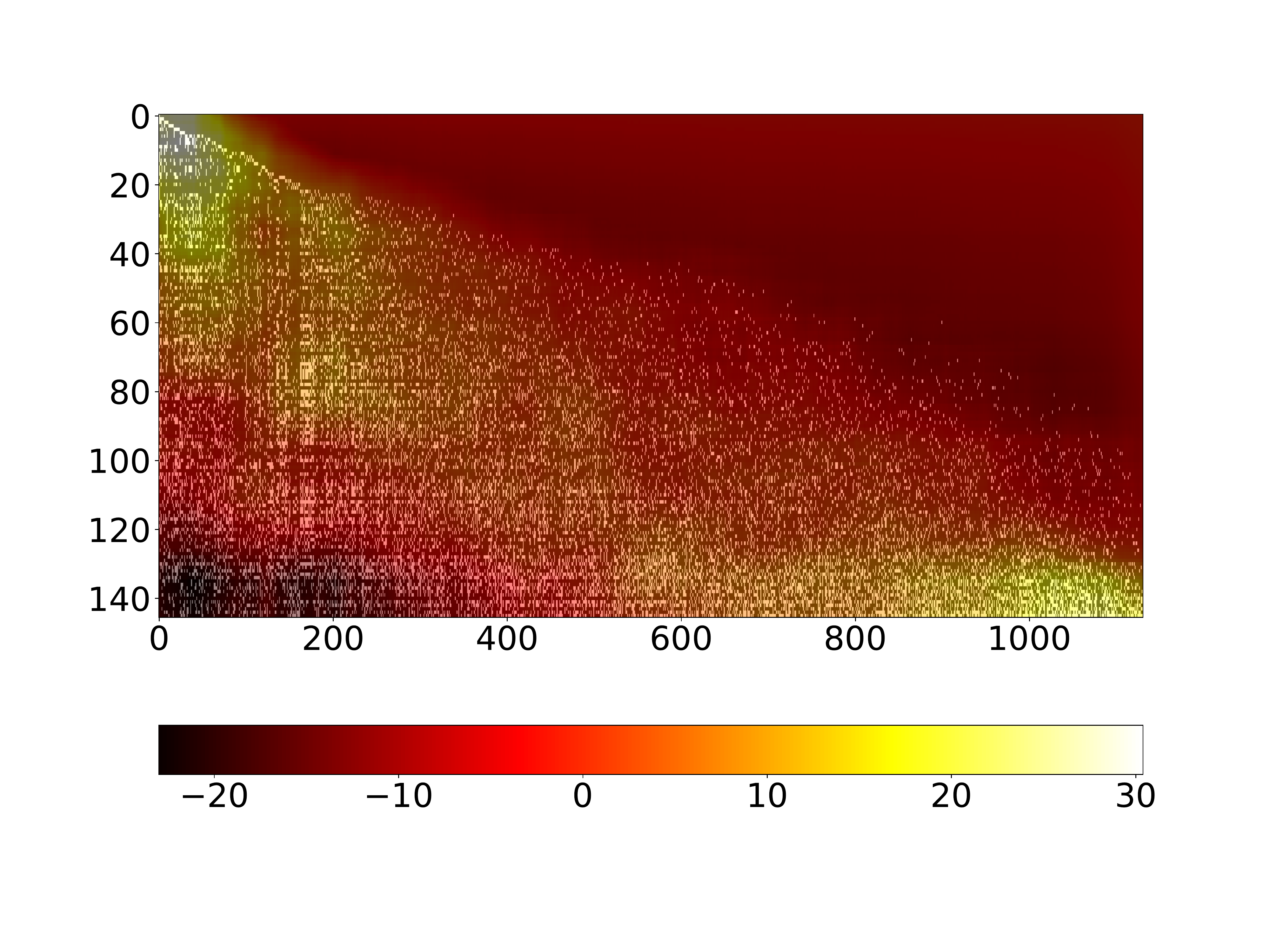}
	\end{center}
	\caption{The images shows the biadjacency matrix in \fig{fig:triangularITN} with links as white dots. The z-scores of the connectivity with respect to the BiCM are shown as a superimposed color map (gray shading).  Higher link abundances than expected are shown as lighter colors, and lower abundances as dark colors. The smaller exporters on the top focus on the most basic products, as shown in the upper-left corner by the z-scores $(z\sim 30)$. The most developed countries on the bottom, on the other hand, specialize on the most sophisticated products, as measured by the scores in the lower-right $(z\sim 25)$. Moreover, they export basic product much less than expected, as can be seen in the the lower-left corner $(z\sim -20)$. This indicates that countries diversify their export baskets as much as possible while specializing on the most sophisticated products they are able to export. Image taken from~\cite{Straka2017}}
	\label{fig:density}
\end{figure*}
\begin{figure*}[ht!]
	\begin{center}
		\includegraphics[width=0.44\textwidth]{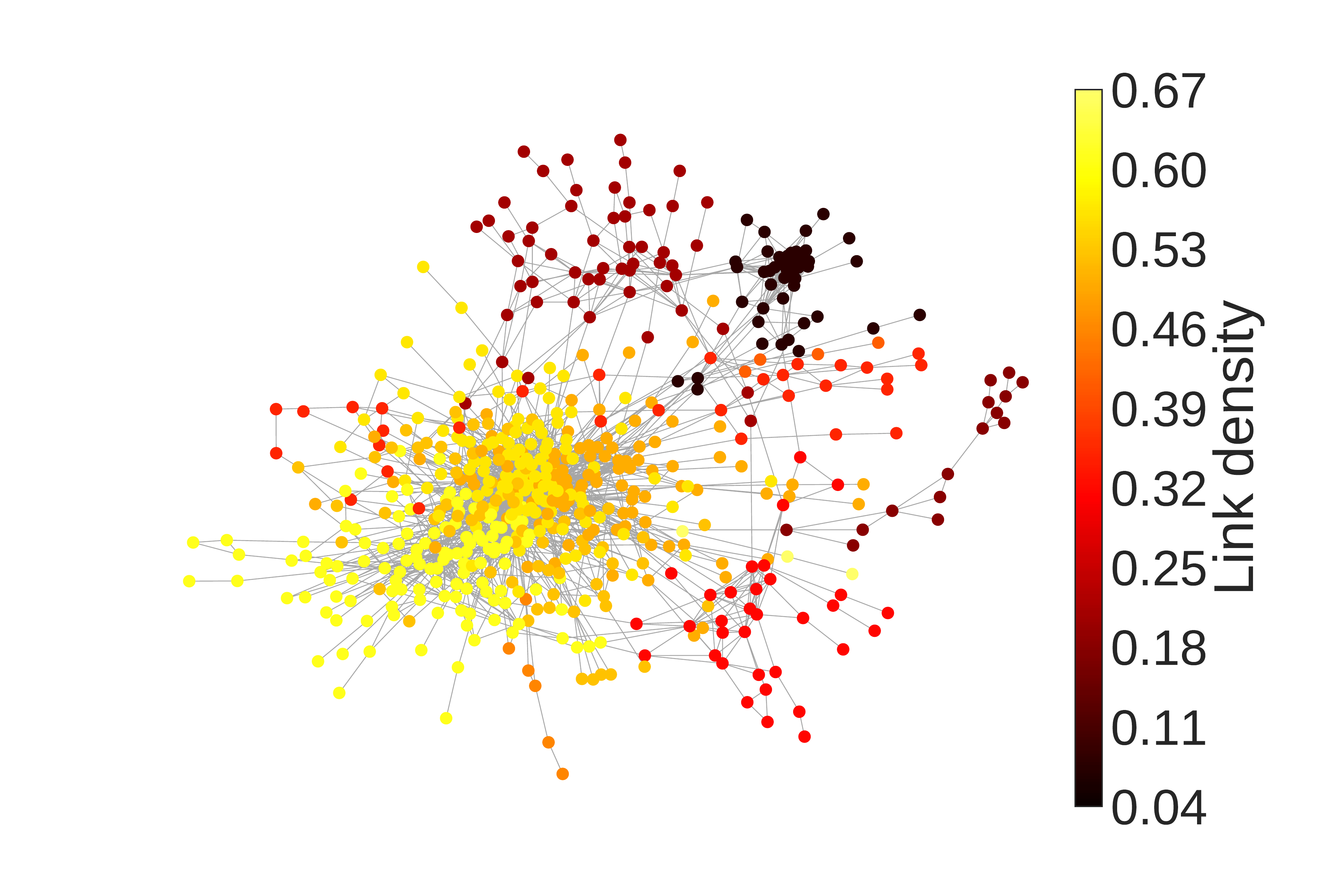}
		\hspace{0.1\textwidth}
		\includegraphics[width=0.44\textwidth]{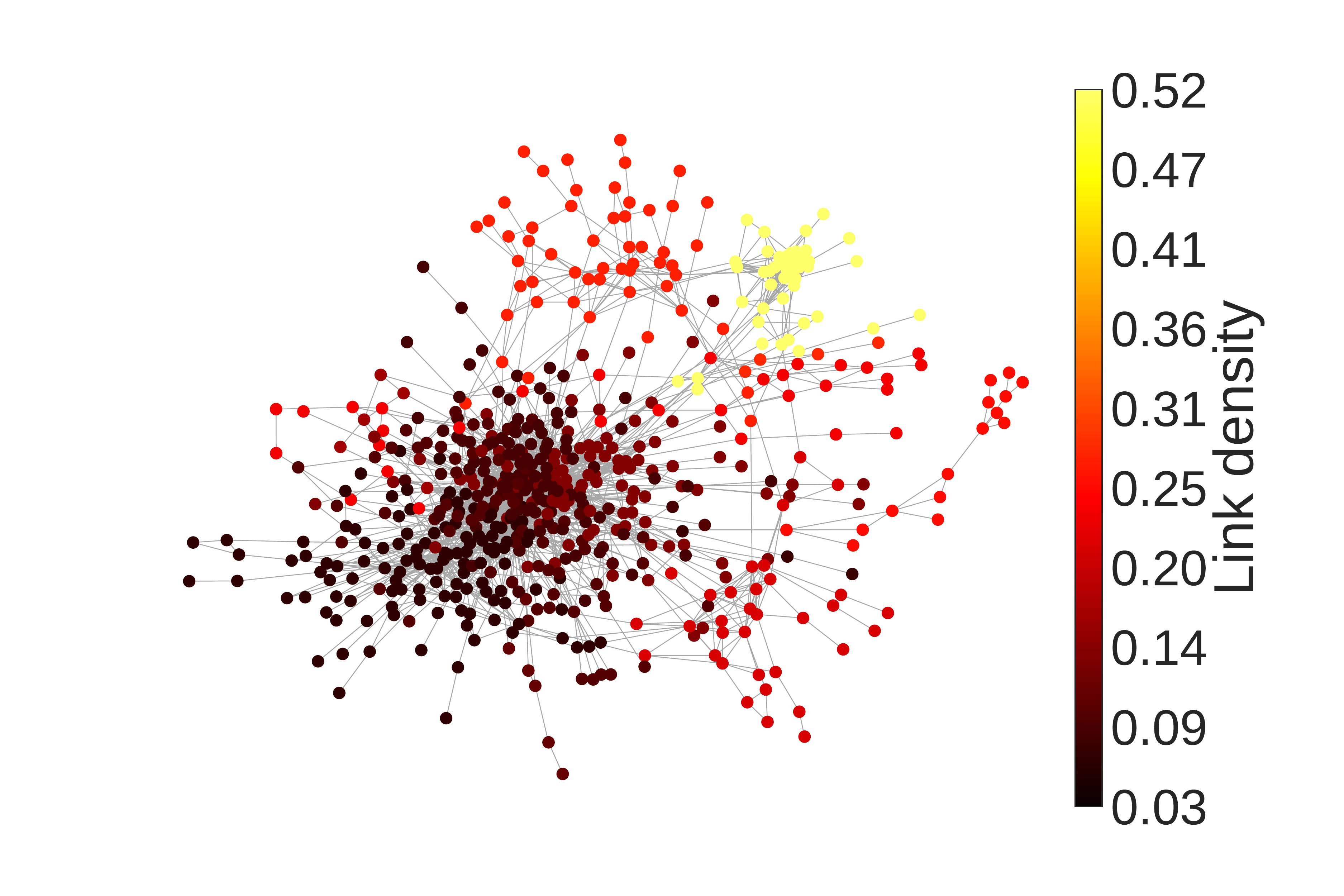}
	\end{center}
	\caption{Illustration of the product network with colors according to the occupation density of country communities~\cite{Straka2017}. \textbf{Left:} Advanced economies occupy the core of the network, which contains high technology items. \textbf{Right:} Developing economies occupy much more the periphery with less sophisticated product rather than the center. Images taken from~\cite{Straka2017}}
	\label{fig:Straka2017_occupations}
\end{figure*}
In~\cite{Tacchella2012,Hidalgo2009,Hausmann2011,Cristelli2013,Hidalgo2007,Caldarelli2012,Zaccaria2014}, is has been observed that the country-product bipartite network is nested, meaning that smaller export baskets are contained in larger ones of more developed nations. For mutualistic networks, it has been argued that the degree distribution may have a strong impact on the observation of nestedness~\cite{Johnson2013}. Using the randomized BiCM, in~\cite{Saracco2015a} this idea has been tested for the trade network and the nestedness of the ITN could be reproduced, thereby underlining the importance of the degree sequence for the nestedness. However, the authors also argue that the BiCM cannot fully reproduce the disassortativity of the ITN, leading to the conclusion that the degree sequence alone is not enough to explain why strong exporters preferably connect to weak exporters. In addition, the numerical quantities fitness and complexity have been compared between the empirical network and the BiCM. As shown in~\cite{Saracco2015a}, both quantities can be reproduced with high precision.
Although the degree information is used only indirectly in the fitness-complexity-algorithm (see, e.g.~\cite{Pugliese2014,Tacchella2012}), this observation underlines that the degree information is enough to account for the fitness and complexity values.

The analysis of the triangular structure of the country-product matrix has been extended in~\cite{Straka2017}. Discounting the degree sequence and comparing empirical with expected link abundances, the authors have observed a specialization signal within the overall export diversification tendency. \fig{fig:density} shows the phenomenon quantitatively using the empirical biadjacency matrix and the z-scores of the expected number of links as a heat map. ``Hotter'' (whitish) colors represent higher z-score values, ``colder'' (dark) colors lower values. The high z-scores stretch from the upper-left to the lower-right and illustrate that countries concentrate their exports on more complex products more than expected, while still exporting basic goods as well. As a consequence, stating that exports baskets are nested is only partially true, since the density in the baskets of more developed countries remains biased towards more exclusive products. 

The previous conclusions can also be drawn from validated monopartite projection of the bipartite trade network (see the section ``Validated Projection''). \fig{fig:Straka2017_occupations} shows the link density between different areas of the product network and advanced economies (left) and developing economies (right):
in both cases, the network is not populated uniformly, since the formers tend to occupy preferentially the highly technological items in the core of the network, whereas the latter focus their export on lower complexity products that belongs to the periphery of the validated product network. In this sense, developed countries, given the size of their export basket, tend to specialize their exports towards the most exclusive products.
\\
\indent The previous result on the biadjacency matrix and the product network therefore reconcile the apparent conflict between Ricardo's argument of export specification and the overall diversification reported in literature~\cite{Strauss-kahn2011}.

\subsubsection{Motif Validation in Trade}

\paragraph{Evolution of Bipartite Motifs of Countries}

In the economic literature, acronyms are often used to refer to groups of
countries that supposedly share similar features in their development and
institutional structures. Famous examples are the G7 (Canada, the
USA, Italy, France, Germany, the UK, Japan), which share a large part of
global, and the five rising BRICS economies (Brazil, Russia, India, China, South Africa). 
Further groups
are, e.g., the MINT countries (Mexico, Indonesia, Nigeria,
Turkey) that show
interesting economic developments~\cite{bloomberg2013} and the south European ``PIGS'' (Portugal, Italy Greece, Spain) that were struggling during the 2008 financial crisis~\cite{Furceri2012}.

Using the bipartite international trade network, it is possible to quantify the similarities within these country groups in terms of their $\text{V}_n$-motifs. In~\cite{Saracco2016a}, the authors compare the real trade network with the randomized ensemble to observe if such similarities are genuine or can just be attributed to the dimension of the export baskets, i.e. the degrees. They applied the BiCM to the product-country trade network and calculate the number of $\text{V}_n$-motifs for each country group, where $n$ is the number of members.
\begin{figure*}
	\begin{center}
		\includegraphics[width=0.7\textwidth]{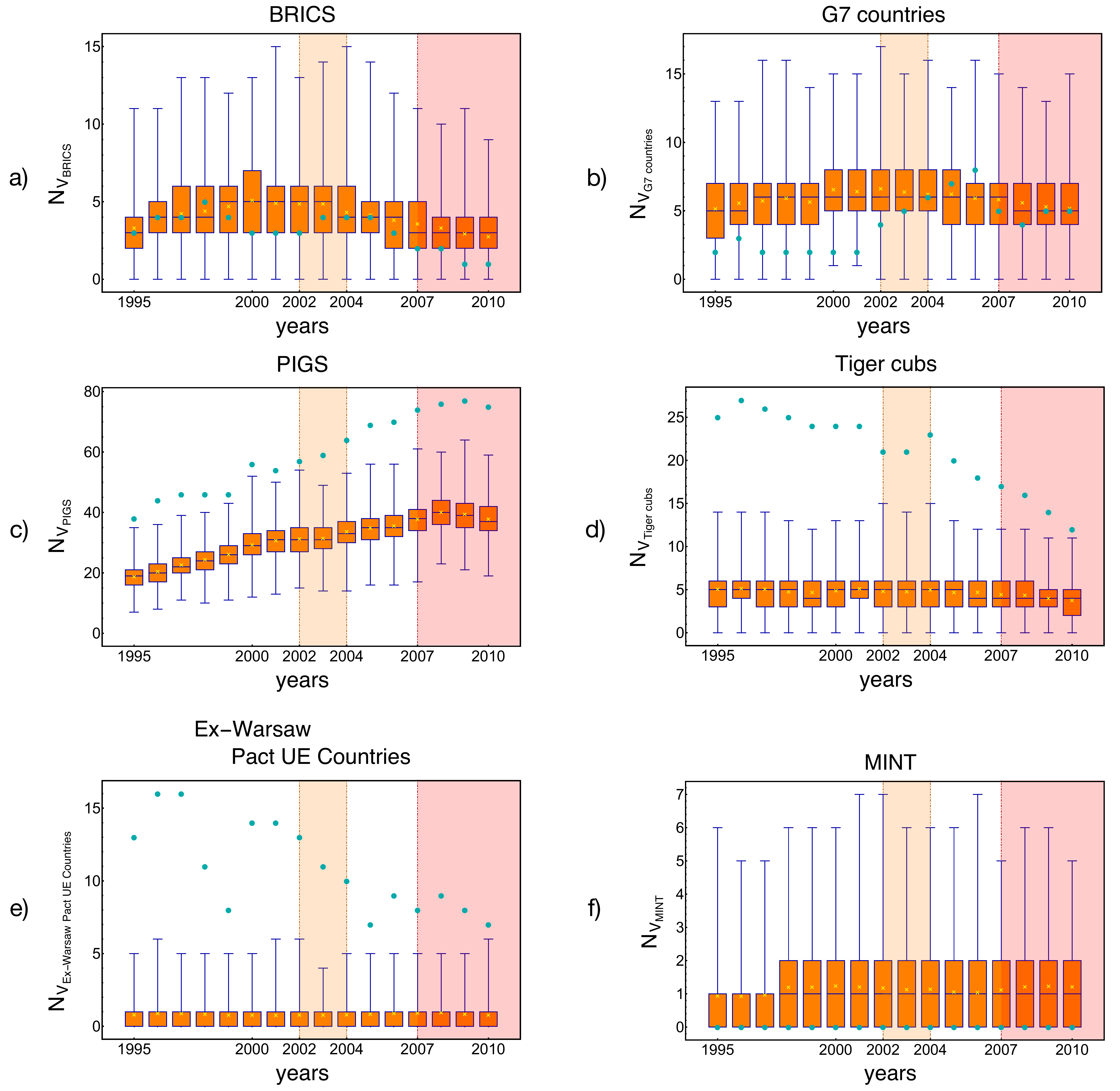}
	\end{center}
	\caption{Comparison of the observed numbers of V$_n$-motifs. The green dots represent the empirical quantity measured on the real network, the box plot the expectation value distribution according to the BiCM. The whiskers capture the 0.15th and the 99.85th percentile. Whereas the BRICS and G7 are compatible with the null model predictions, PIGS, Tiger Cubs and the ex-Warsaw Pact countries show abundances that are not explainable in terms of the degree sequences alone. Image taken from~\cite{Saracco2016a}.}
	\label{fig:countries}
\end{figure*}

\fig{fig:countries} compares the number of $\text{V}_n$-motifs
for different country
groups~\cite{Saracco2016a}: green dots represent the values observed in the
data, whereas the box-plots capture the probability distribution of the
ensemble. In panel c) we can see that the observations for the PIGS 
lie clearly above the box-plot whiskers, which indicates that the similarities is not merely due to their export baskets sizes. This is true even before the arrival of the 2008 crisis, although the discrepancy
gradually increases over time. Contrary to the PIGS, the BRICS  in panel a)
show export basket overlaps that are compatible with the null model. The MINT countries, on the other hand, do not
have a single export item in common, as shown in panel f). As a consequence,
both MINT and BRICS groupings cannot be justified by the observation of similar
industrial capabilities alone (see also \cite{Tacchella2012,Cristelli2013}).

Contrary to that, strong similarities can be observed in the \emph{Tiger Cubs} (Thailand, Indonesia,
Malaysia, Philippines, Vietnam), which experienced a recent industrialization process
similar to the original \emph{Four Asian Tigers} (Hong Kong, Singapore, South Korea,
Taiwan). Panel d) in \fig{fig:countries} shows a statistically
significant signal of V$_n$-motifs, which gradually diminishes in intensity. This
indicates that their recent industrial developments began to diverge,
progressively turning into a differentiation in their exports.
Similarly, the impact of a common communist industrialization program can be
observed in the exports of \emph{ex-Warsaw Pact countries} that are now part of the European Union (such as Poland, Romania,
and Hungary) well into the years 2000, see panel e) in
Fig.~\reff{fig:countries}. After joining the EU, the signal has progressively declined. The composition of the G7 group, on the other hand, can be simply attributed to their degrees, i.e. to the dimensions of their export baskets. Panel b) shows that not statistically significant signal can be detected.

\paragraph{Closed Motifs evolution} 
Closed motifs are more complex combinations of links and capture mesoscopic structural properties of the network. For instance, a X-motif measures how many times two countries compete on the
world market by sharing more than one product in their export basket. A high number of
X-motifs indicates that their export baskets are very similar on a global
level. Several scholars~\cite{WorldEconomicForum2013} suggest that an excessive degree of similarity in
industrialization and exportation weakens the international trade network
and makes is more prone to stress. A diversification of industrial
capabilities, on the other hand, would make the system more resilient.

Following the trends in panels a), b) and c) of Fig.~\reff{fig:motifs}, we can
observe an increase in the number of closed motifs, i.e. of the similarity of
export baskets, before the financial crisis~\cite{Saracco2016a}. This development comes to an
abrupt halt in 2008 and abundances drop from that point on. Notice that this
evolution only illustrates what happened in the wake of the crisis and does
not provide any early-warning signal. However, comparing the observations with
the null model, a very different picture arises, as illustrated in panel d): all three motifs occur
less in the network than would be expected from the BiCM. Z-scores are
predominantly negative, with values as small as -3 to -4 for the X- and
M-motifs. They are relatively stable until 2003, from which point on we observe
a clear trend towards greater z-scores, i.e. towards a better agreement with
the null model. Thus when the crisis struck in 2008, the international trade network had already undergone significant structural changes in the preceding five years, which eventually fade out around 2010.

Note that the last results do not imply any causal relation, as discussed in~\cite{Saracco2016a}. 
However, the dramatic evolution of the closed motifs anticipated the crisis. In that sense it can be regarded as an early-warning signal, since it informs us that the network is changing globally such that the resilience of the whole system is modified.

Moreover, it is worth underlining that these observations cannot be made without the application of the null model. Considering only the data limits us to the simple observation of abrupt trend changes in motif abundances. Only the comparison with the BiCM reveals a clear signal of anticipating structural change. Analyses as the one presented here provide effective tools to monitor bipartite network, such as the ITN, and offer deep insight into the state of the system to policy makers, motivating them to act instead of react.

\begin{figure*}[t]
	\begin{center}
		\includegraphics[width=0.7\textwidth]{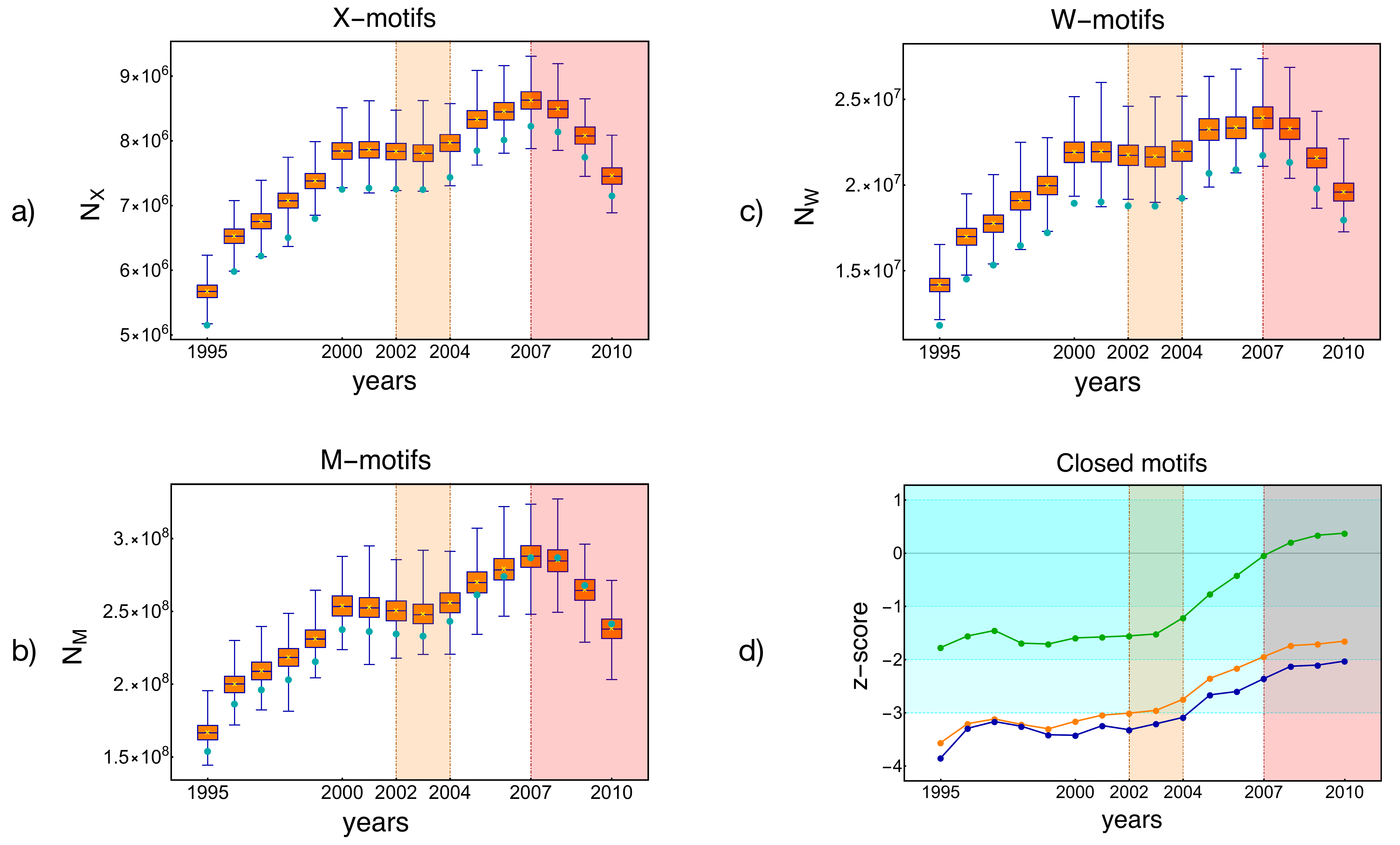}
	\end{center}
	\caption{Panels a), b), and c) show the abundance of closed motifs in the ITN network from 1995 until 2010. It is apparent that they increased until the financial crisis struck and dropped after 2007. The values are compared with the BiCM expectations in panel d) using z-scores. Before the crisis, abundances were strongly underrepresented and increasing significantly already from 2003 onwards, indicating a global change of the network. Image taken from~\cite{Saracco2016a}}
	\label{fig:motifs}
\end{figure*}

In summary, the recent results on the international trade network obtained through the comparison with the BiCM has shown that many apparently genuine properties can be traced back to the degree sequence, i.e. the number of exporters of a product and the length of the export basket of a country.

\section{Financial Networks}
Financial institutions form a global system of investments and money lending. In the aftermath of the 2008 financial crisis, correctly assessing systemic risk and shock propagation has become a top priority for policy makers and regulators. Contrary to previous beliefs, the financial network has revealed itself to be more unstable than expected due to the complex structure of the connections~\cite{ChanLau2009,Brunnermeier2009,Krause2012,Arinaminpathy2012,Battiston2016}.

Financial stress can be transmitted through two main channels: direct exposure due to bilateral agreements, such as credit swap contracts~\cite{Greenwood2015}, and indirect exposure due to portfolio overlaps~\cite{Allen2000,Eisenberg2001,Gai2010}. Whereas the first gives rise to an inter-bank network, the second presents itself naturally as a bipartite network.

Interest in the inter-bank network has surged in the fields of public administration and academic research ever since the bankruptcy of Lehman Brothers and the subsequent turmoil. An important contribution of network theory has been to shift the paradigm from the dogma ``too big to fail'' to ``too central to fail''~\cite{Battiston2012}. To quantify the financial risk associated to different institutions including network effects, the so-called ``DeptRank'' was introduced in~\cite{Battiston2012}.
 
Indirect exposure, on the other hand, can be created through bank portfolio overlaps. In a bank-asset network, financial institutions are ordered along one layer and assets (or asset classes) along the other. Financial contagion can be created through \emph{fire sales} spillover effects: a sudden drop in the value of an asset can trigger a cascade of sell-orders, which leads to asset illiquidity~\cite{Shleifer2011,Caccioli2014,Cont2016,Gualdi2016,Greenwood2015,Squartini2017}. This effect can put banks into distress, who may react by selling other assets, thereby causing further devaluation dynamics.

In an recent article, a dynamical model for the analysis of shocks in the bank-asset network has been presented and applied to the Venezuelan banking system~\cite{LevyCarciente2015}. The authors show that their model is able to capture temporal changes in the structure of the network and that some assets with small capitalization can cause significant global shocks~\cite{LevyCarciente2015}.  Fire sale spillovers have also been analyzed by~\cite{Greenwood2015}, who have introduced a metric to asses the systemic risk of the bank-asset network.

Despite these significant advancements, the analysis of financial network is often hindered by a lack of detailed data. The model in~\cite{LevyCarciente2015}, for instance, uses balance sheets for the model construction -- but often, such information is available only in aggregate and detailed asset holdings are undisclosed. Many tools of financial analysis therefore rely on aggregate data, resulting in unrealistically dense networks and a biased underestimation of systemic risk~\cite{Squartini2017}. As a consequence, improved methods are necessary that reconstructed the network in a more realistic way while avoiding systematic bias~\cite{Squartini2017}.

\subsection[]{Bipartite Exponential Random Graph III. -- Weighted Networks}

We have already seen that the bipartite exponential graph can be used to construct unbiased statistical benchmark models. The BERG framework can be easily extended from binary to weighted networks.

In weighted bipartite networks, nodes are characterized by their degrees and strengths, i.e. the sum over the weight of their edges. If only the node strengths are available, for instance in the case of aggregate portfolio positions of banks, one may intuitively be inclined to extend the BiCM to its weighted counterparts, the \emph{bipartite weighted configuration model} (BiWCM~\cite{diGangi2015}, see Appendix~\reff{sec:appendixNullModels}), by simply exchanging the degree with strength constraints. However, it has been shown for monopartite networks that the reconstruction of such network performs very badly~\cite{Mastrandrea2014}. This is due to the fact that it ignores the information on the network topology that is contained in the binary degree sequence. In fact, the BiWCM has shown to seriously underestimate risk exposures in the bank-asset bipartite network~\cite{diGangi2015}. As the authors of~\cite{Mastrandrea2014} point out, non-trivial degree and strength sequences complement each other in the network reconstruction. The constraints should thus be modified accordingly.

\begin{figure}[t]
	\begin{center}
		\includegraphics[width=0.7\textwidth]{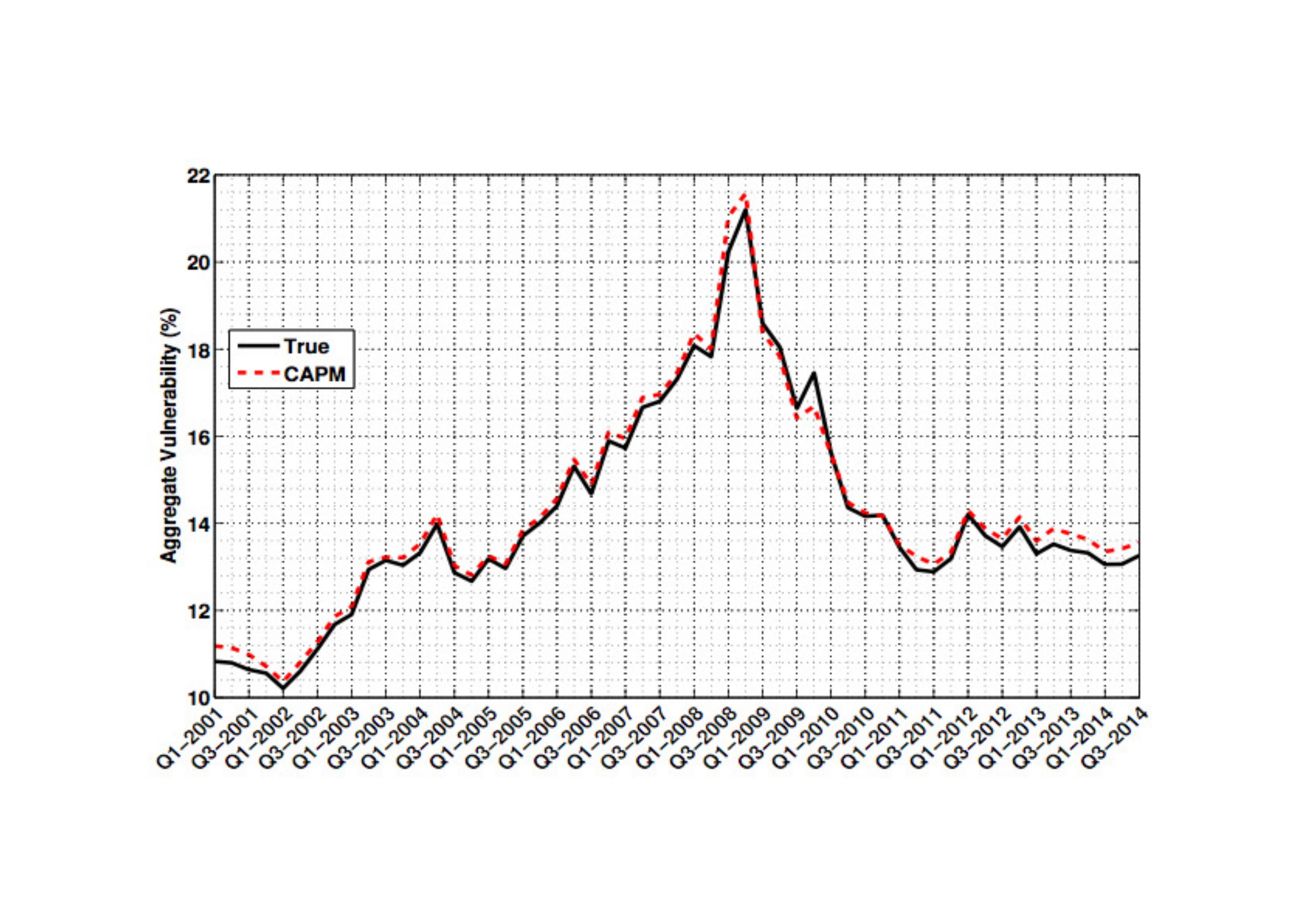}
	\end{center}
	\caption{Aggregate vulnerability of the bank-asset network calculated with the metric calculated by Greenwood et al.~\cite{Greenwood2015} using the whole data (solid black line) and the CAPM matrix weights, which require only the node strengths (dotted red line). Image taken from~\cite{diGangi2015}.}
	\label{fig:diGangi2015CAPM}
\end{figure}

Given the weighted biadjacency matrix $\bold{W}$, we can obtain the node strengths by summing over the rows and columns, respectively. Let us index the banks with $i\in\Gamma$ and the assets with $\alpha\in\text{L}$. In the financial context, the vertex strengths are often described as the \emph{total asset size} of a bank (or \emph{market value} of their portfolio), $V_i = \sum_{\alpha} w_{i\alpha}$, and the \emph{market capitalization} of an asset, $C_\alpha = \sum_{i} w_{i\alpha}$~\cite{diGangi2015,Squartini2017}. It is possible to remap the matrix entries in such a way that banks choose their portfolio weights proportional to their market value and the asset's capitalization:
\begin{equation}\label{eq:wCAPM}
w^{CAPM}_{i\alpha} = \frac{V_i C_\alpha}{w},
\end{equation}
where we have used $w = \sum_{i', \alpha'} w_{i'\alpha'}$
This model is called \emph{capital asset pricing model} (CAPM,~\cite{Lintner1965,Mossin1966}). As has been shown in~\cite{diGangi2015} and is illustrated in \fig{fig:diGangi2015CAPM}, these matrix weights give a good approximation of the systemic risk of the system measured in terms of the metric introduced by~\cite{Greenwood2015}, despite the fact that networks of return price correlations show little agreement with real cases~\cite{Bonanno2003,Bonanno2004}. However, without the use of a null model little can be said about the precision of the risk predictions~\cite{diGangi2015}.

\subsection{Systemic Risk}
\begin{figure*}[t]
	\begin{center}
		\includegraphics[width=0.6\textwidth]{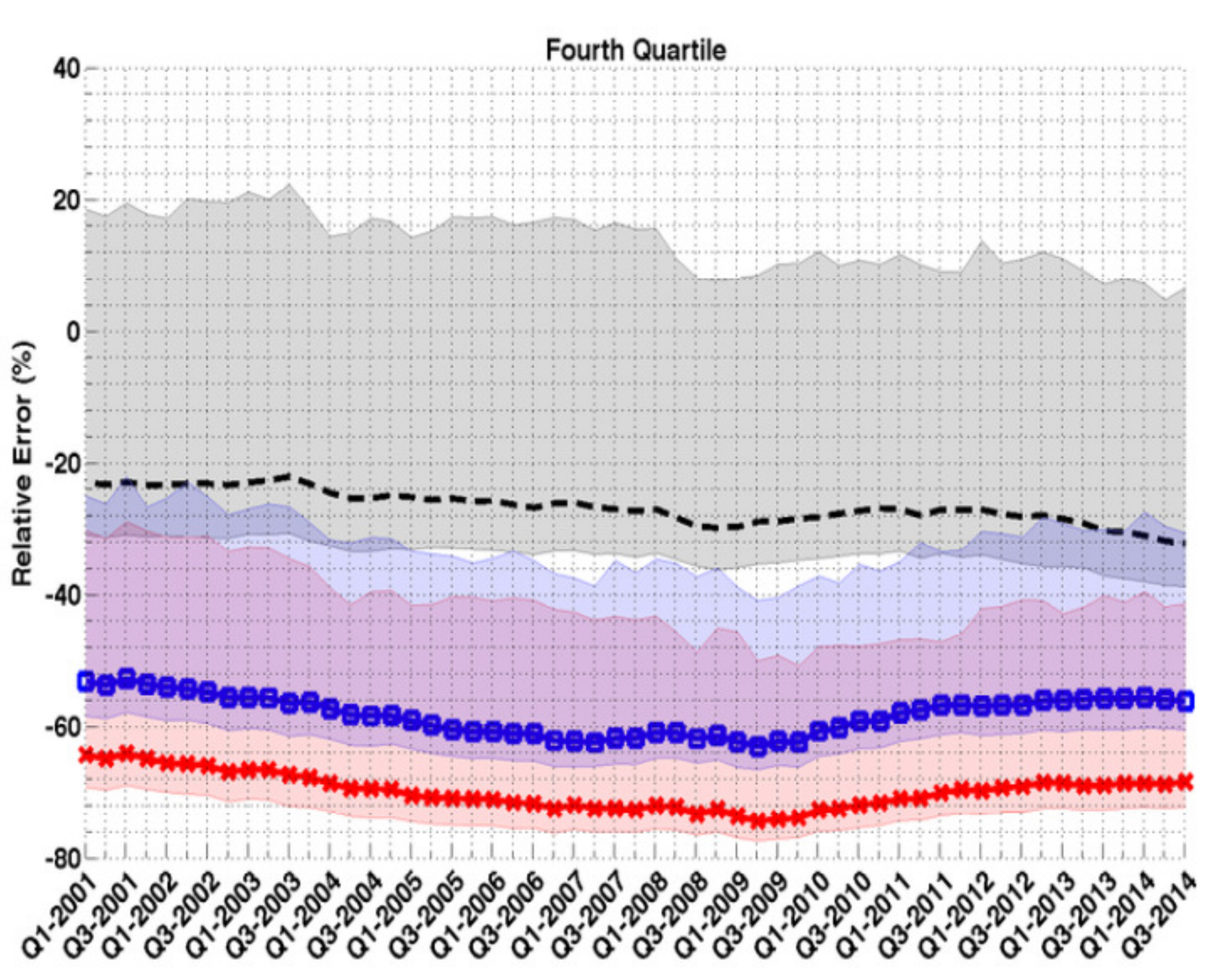}
	\end{center}
	\caption{Quartile of banks with the highest indirect risk as measured by~\cite{diGangi2015} during the interval 2001 -- 2014. \textbf{MECAPM:} dashed line, grey shading; \textbf{BiECM:} blue squares, blue shading; \textbf{BiWCM:} red crosses, red shading. Although all three models systematically underestimate the systemic risk, MECAPM outperforms the best. Image taken from \cite{diGangi2015}.}
	\label{fig:diGangi2015Systematicness}
\end{figure*}
In order to assess the performance of benchmark models in estimating systemic risk, in~\cite{diGangi2015} fire sales spillover effect have been considered on the bank-asset network of US commercial banks. Their data is derived from quarterly reports which disclose the single positions in the bank portfolios. Hence, the authors could compare the risk estimations due to aggregate exposures, considering only the node strengths, with measures that take also the degrees into account. Risk is measured using the metric defined by Greenwood et al.~\cite{Greenwood2015}.

To construct an null model that reflects the risk estimation of the CAPM framework, the  \emph{maximum entropy capital asset pricing model} (MECAPM,~\cite{diGangi2015}, see Appendix~\reff{sec:appendixNullModels}) has been defined, which respects the constraints $\langle w_{i\alpha}\rangle = w^{CAPM}_{i\alpha}$. Notice that, while the BiWCM imposes $N_\text{L} + N_\Gamma$ constraints, the MECAPM uses $N_\text{L} \times N_\Gamma$ conditions. 

The probability distribution for the MECAPM yields~\cite{diGangi2015}
\begin{equation}
P(G_B) = \prod_{i, \alpha} \left[1 - \left(p_{CAPM}\right)_{i\alpha}\right]^{w_{i\alpha}} \left(p_{CAPM}\right)_{i\alpha},
\end{equation}
where the probability per link reads~\cite{diGangi2015,Squartini2017}
\begin{equation}
\left(p_{CAPM}\right)_{i\alpha} = \frac{w^{CAPM}_{i\alpha}}{1 + w^{CAPM}_{i\alpha}}.
\label{eq:pMECAPM}
\end{equation}
$P(G_B)$ is thus geometrically distributed~\cite{diGangi2015}. 

Since strength and degree information can complement each other, in analogy to the monopartite case in~\cite{Mastrandrea2014} the authors of~\cite{diGangi2015} have also included the so-called \emph{Bipartite Enhanced Configuration Model} (BiECM, \cite{diGangi2015}), on which degrees and strength constraints are 
imposed. As shown in~\cite{diGangi2015} and summarized in Appendix~\reff{sec:appendixNullModels}, the graph probability yields 
\begin{equation}
P(G_B|\bold{W}) = \prod_{i, \alpha} \frac{(1 - \phi_i \xi_\alpha) (\phi_i\xi_\alpha)^{w_{i\alpha}}(\psi_i\gamma_\alpha)^{\Theta(w_{i\alpha})}}{1 - \phi_i \xi_\alpha(1 - \psi_i\gamma_\alpha)},
\end{equation}
where we have used the short-hand notation $\phi_i = e^{-\rho_i}, \xi_\alpha = e^{-\rho_\alpha}, \psi_i = e^{-\theta_i}$ and $\gamma_\alpha = e^{-\theta_\alpha}$, and $\theta$ and $\rho$ are the Lagrange multipliers for the degrees and strengths, respectively.

The results of the analysis are summarized in \fig{fig:diGangi2015Systematicness} for the banks with the highest systemic exposures in the data. Although all three models systematically underestimate risk, MECAPM clearly outperforms the other two models~\cite{diGangi2015}. Notice that the BiWCM performs very badly, underestimating the risk as much as -80\%. Errors are relatively large, as we can see in the shaded areas. \\

A possible reason for the large error intervals in \fig{fig:diGangi2015Systematicness} has been suggested in~\cite{Squartini2017}, pointing to the fact that MECAPM predicts very dense network configurations. In fact, from ~\eq{eq:pMECAPM} we can see that the link probabilities quickly approach 1 for $w_{i\alpha}^{CAPM} \gg 1$. This issue has been taken up in~\cite{Squartini2017}, who formulate the so-called \emph{Enhanced Capital Asset Pricing Model} (ECAPM,~\cite{Squartini2017}). Using only the strength sequences $C_i$ and $V_\alpha$, this approach aim at reconstructing the network topology while imposing the CAPM link weights.

Firstly, the topology of the network is established by using the BiCM under the assumption that the Lagrange multipliers are proportional to node-specific fitness values, represented by their strengths~\cite{Squartini2017}. In analogy to the BiCM (see Appendix~\reff{sec:appendixNullModels}), this gives us
\begin{equation}
(p_{ECAPM})_{i\alpha} = \frac{z V_i C_\alpha}{1 + z V_i C_\alpha},\quad \forall i \in \Gamma, \alpha \in L,
\end{equation}
where $z$ absorbs the proportionality constants.

Secondly, the link weights are reconstructed using the CAPM model while taking the network topology into consideration. Instead of setting $w_{i\alpha} = V_i C_\alpha / w$, a correction factor is applied~\cite{Squartini2017}
\begin{equation}
\begin{split}
w_{i\alpha} 
&= m_{i\alpha} \frac{V_i C_\alpha}{w\ (p_{ECAPM})_{i\alpha}} \\
&= (V_i C_\alpha + z^{-1}) \frac{m_{i\alpha}}{w},
\end{split}
\end{equation}
where $m_{i\alpha}$ is 0 or 1 depending on the link. As pointed out in~\cite{Squartini2017}, the weight expectations of the ECAPM correspond with those of the MECAPM and CAPM~\cite{Squartini2017}. However, the former reconstructs the network topology separately, which compensates the high network densities for the latter.

\begin{figure*}[t]
	\begin{center}
		\includegraphics[width=0.9\textwidth]{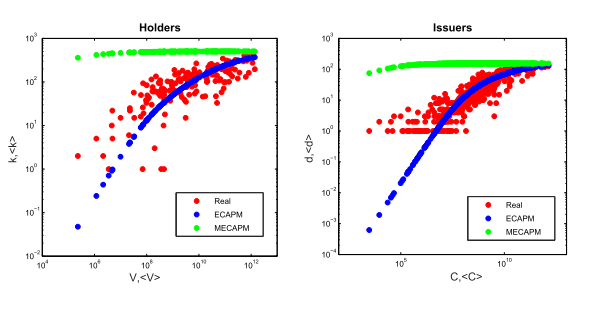}
	\end{center}
	\caption{Relation between the strengths and degrees of the nodes in~\cite{Squartini2017}. The security holder portfolios have certain market values $V_i$ (the ``banks'') and issued security assets have a market capitalization $C_\alpha$ (the ``assets''). The ECAPM follows the real empirical data much closer than the MECAPM, which systematically overestimates the degrees. Image taken from~\cite{Squartini2017}.}
	\label{fig:squartini2017StrengthDegree}
\end{figure*}

The difference between the ECAPM and the MECAPM has been tested in~\cite{Squartini2017} on a data set of security holdings of European institutions (Security Holding Statistics, SHS) collected by the European Central Bank from 2009 to 2015.
The empirical difference is visible in~\fig{fig:squartini2017StrengthDegree}, which compares the node degrees with node strengths. The MECAPM shows continuously high degrees and does not capture the real distribution, which illustrates the observation made for \eq{eq:pMECAPM}. Even though the ECAPM underestimates degrees for small strength values, it reproduces the data better.

The since both, MECAPM and ECAPM, reproduce the same weights, they estimate the same systemic risks as measured with the metric introduced in~\cite{Greenwood2015}. However, reconstructing the topology as in~\cite{Squartini2017} significantly decreases the uncertainty of the risk metric. In particular, comparing the errors of the MECAPM and ECAPM yields 
\begin{equation}
\sigma_{S_i}^{MECAPM} \propto V_i^{1/2}\sigma_{S_i}^{ECAPM},
\end{equation}
where $V_i$ is the value of institution $i$ and $S_i$ its systematicness~\cite{Squartini2017}. The fluctuations in the ECAPM framework are thus systematically smaller than in MECAPM and motivate the application of a degree as well as strength reconstruction.

\subsection{Portfolio Overlap Projection}
\begin{figure}[ht]
	\begin{center}
		\includegraphics[width=0.45\textwidth]{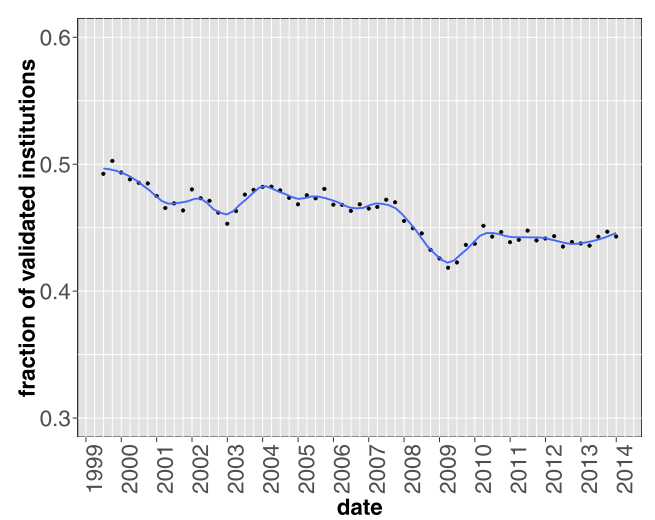}
		\hspace{0.08\textwidth}
		\includegraphics[width=0.45\textwidth]{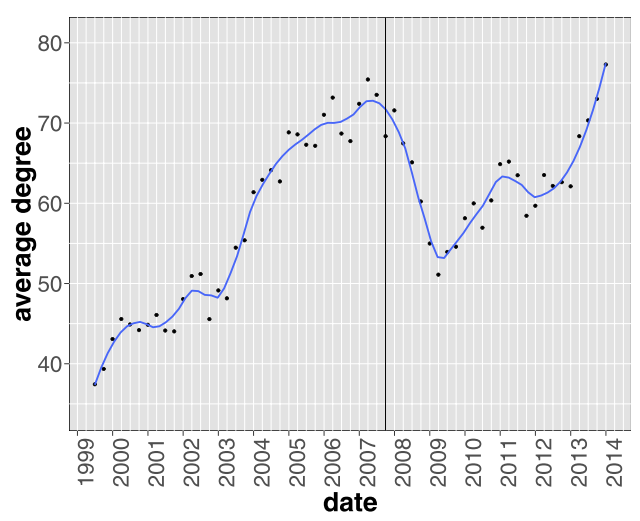}
	\end{center}
	\caption{\textbf{Left:} Fraction of financial institutions in the validated monopartite projection. The values decrease only slightly, with temporary reduction around 2009. \textbf{Right:} Average degree in the monopartite projection of financial institutions. Notice the salient dip in 2009 after the beginning of the crisis. Images taken from~\cite{Gualdi2016}.}
	\label{fig:gualdi2016}
\end{figure}
The risk of fire sales has also been treated in~\cite{Gualdi2016} using a different methodology. Contrary to~\cite{diGangi2015} and~\cite{Squartini2017}, they do not consider weighted networks but instead the binary bipartite network between financial institutions and their asset holding in the years 1991 -- 2013. Instead of calculating the systemic risk measure on the null model, they focus on the overlap matrix of portfolios, which expresses the number of assets that financial institutions share, i.e. the number of their V-motifs (see \eq{eq:V}). By applying the grand canoncical projection algorithm~\cite{Saracco2016} with a Bonferroni correction for the p-value testing, and thus comparing the observed values V$^*$ with their BiCM expectations and validating only statistically significant links, they obtain a network of financial institutions containing only relevant links that are not accounted for by the degree sequences.

As \fig{fig:gualdi2016} shows, the fraction of institutions that have at least one significant edge remains relatively constant. In spite of this seemingly innocuous development, the similarity of these nodes increase very quickly, as illustrated in \fig{fig:gualdi2016}. In particular, notice how the similarity increases before the 2007--2008 financial crisis. After a drop in 2009, it took up pace and has reached levels even higher than before the crisis~\cite{Gualdi2016}.
 
In conclusion, the authors point out that the validated projection method can recover those financial institutions that would be at risk of suffering the greatest losses in cases of financial distress~\cite{Gualdi2016}.

\section{Conclusions and Outlook}
In this review-like paper we have revised a number of techniques, designed for the analysis of bipartite networks. Interestingly, while many of the quantities describing bipartite networks have been defined within the field of ecology, several techniques that have been developed afterwards to test the statistical significance of the same quantities were born in a - seemingly - completely different context, e.g. in economic and financial systems. However, while the study of the latter has benefited from the empirical observations carried out in the former,  many of the analytical tools commonly employed to study socio-economic systems seem not to have been recognized yet as potentially useful for ecological systems. One of the aims of this work is precisely that of bridging the gap between two - apparently - distant fields which can greatly benefit from the advances of each other.

A straightforward example is provided by the detection of mesoscale structures, as communities and motifs. While the techniques that have been recently proposed to overcome the limitations of modularity maximization have not crossed the border of graph theory, their application to the study of ecological networks is still in its embrional stage~\cite{Baiser2016}. The same holds true for what concerns the application of the multiplex formalism~\cite{Pilosof2017}. The importance of such a topic is clearly shown by the analysis of social networks as ``mutualistic'' information ecosystems, where it has been recently found that mesoscale structures may be correlated to the emergence of collective attention~\cite{Borge-holthoefer2017}, in particular when a transition from a modular to a nested structure is observed. The aforementioned results also point out the need to develop dynamical models to study the evolution of bipartite networks.

An application that has benefited from both the concepts developed within ecology and the methods developed within the field of social networks is represented by the so-called recommendation systems~\cite{Zhou2007}. Briefly speaking, the latter are algorithm intended to suggest users their next ``choices'', be they items, movies, etc. Although many recommendation algorithms exist, an interesting example is provided by those ones building upon the idea of a \emph{resource-allocation} dynamics taking place on the network. Other concepts such as \emph{specialization} and \emph{interaction} have inspired models able to reproduce observed patterns in both ecological and social systems~\cite{Saavedra2011}.

An even more recent advancement is represented by the employment of \emph{tripartite} networks to study the relationships between technology and economic development~\cite{Pugliese2017}. The three layers considered there are represented by technologies, countries and products and the analysis aims at quantifying the probability of jumping from a given technology in layer 1 to a given product in layer 2, while accounting for all possible paths through the intermediate countries layer. Although the null model employed for the analysis is, actually, the combination of two distinct BiCMs, the paper represents an interesting future direction of research.

\section*{ACKNOWLEDGMENTS}
This work was supported by the EU projects CoeGSS (Grant No. 676547), (Grant No. 317532), Openmaker (Grant No. 687941), SoBigData (Grant No. 654024), and the FET projects SIMPOL (Grant No. 610704), DOLFINS (Grant No. 640772).

\appendix
\section{Appendix: Revealed Comparative Advantage}\label{sec:appendixRCA}
The \emph{revealed comparative advantage} (RCA, also knows as Balassa index~\cite{balassa1965trade}), rescales the product export volumes in order to determine whether countries are relevant exporters of products. Be $e(c, p)$ the export value of product $p$ in country $c$'s export basket. The RCA is calculated by comparing the monetary importance of $p$ in $c$'s export basket to the global average, 
\begin{equation}
\text{RCA}_{c, p} = \frac{e(c, p)}{\sum_{p'} e(c, p')} \left/ \frac{\sum_{c'}e(c', p)}{\sum_{c', p'} e(c', p')} \right. .
\end{equation}
A country is a relative exporter if RCA $\geq 1$. Using the RCA, the weighted country-product biadjacency matrix can be binarized by keeping only those matrix entries that identify relevant exports and setting them to 1.

\section{Appendix: Bipartite Exponential Random Graph Model}\label{sec:appendixNullModels}
We report some of the null models that have been obtained through maximum entropy maximization and have been applied to binary and weighted bipartite networks. In the following, all quantities marked with an asterisk refer to the real networks, expressed by their binary ($\bold{M}^*$) or weighted ($\bold{W}^*$) biadjacency matrix. The layer dimensions are $N_\text{L}$ and $N_\Gamma$. 

\subsection{Bipartite Random Graph}
Constraining the expected number of links in the graph ensemble yields an extension of the Erd{\H o}s-R{\'e}nyi random graph to bipartite networks, the \emph{Bipartite Random Graph} (BiRG). The constraint $C \equiv E = \sum_{i,\alpha}m_{i\alpha}$, and thus the Lagrange multiplier $\theta$ as well, is scalar. The partition function can be calculated easily: 
\begin{equation}\label{eq:ZRG}
\begin{split}
\mathcal{Z}_\text{BiRG}(\theta)=&\sum_{G_\text{B}\in\mathcal{G}_\text{B}}e^{-\theta E(G_\text{B})}\\
=&(1+e^{-\theta})^{N_\text{L}N_\Gamma}.
\end{split}
\end{equation}
The probability per graph reads
\begin{equation}\label{eq:PBiRG}
\begin{split}
P(G_G|\theta)=&\dfrac{e^{-\theta E}}{(1+e^{-\theta})^{N_\text{L}N_\Gamma}}\\
=&\left(p_\text{BiRG}\right)^E(1-p_\text{BiRG})^{N_\text{E}N_\Gamma-E},
\end{split}
\end{equation}
where $p_\text{BiRG} \equiv \dfrac{e^{-\theta}}{1+e^{-\theta}}$ is the
probability of observing a bipartite link between any node couple $i\in \Gamma$, $\alpha \in \text{L}$. Notice that $p_\text{BiRG}$ is uniform and independent of the links. Since \eq{eq:PBiRG} is a Binomial distribution, we see that the
probability of observing a generic graph $G_\text{B}$ in the ensemble reduces
to the problem of observing $E(G_\text{B})$ successful trials with the same probability $p_\text{BiRG}$. We can obtain an analytical
expression for the Lagrange multiplier $\theta$ and thus for the link probability by maximizing the likelihood, which reads
\begin{equation}
\mathcal{L}=\ln P(G^*|\theta)=-\theta\, E^*- N_LN_\Gamma\ln(1+e^{-\theta}),
\end{equation}
and returns 
\begin{equation}
p_\text{BiRG}=\frac{E^*}{N_L\,N_\Gamma}.
\end{equation}

\subsection{Bipartite Partial Configuration Model} 
Without loss of generality, we constrain the degree sequence on the layer $\Gamma$ such that $\langle
k_i\rangle=k_i^*,\,\forall i \in \Gamma$. For each node degree $k_i$, we have introduce one associated Lagrange multiplier, $\theta_i$. This gives us the \emph{Bipartite Partial Configuration Model} (BiPCM, \cite{Saracco2016}). Following the same procedure as in Eq. \reff{eq:ZRG}, we can obtain
\begin{equation}
\mathcal{Z}_\text{BiPCM}(\vec{\theta})=\prod_{i,\alpha}1+e^{-\theta_i}.
\end{equation}
The probability per graph reads
\begin{equation}\label{eq:PBiPCM}
\begin{split}
P(G_B|\vec{\theta})
=&\prod_{i,\alpha}(p_\text{BiPCM})_i^{m_{i\alpha}}\big(1-(p_\text{BiPCM})_i\big)^{1-m_{i\alpha}}\\
=&\prod_{i}(p_\text{BiPCM})_i^{k_i}\big(1-(p_\text{BiPCM})_i\big)^{N_L - k_i},
\end{split}
\end{equation}
where $(p_\text{BiPCM})_i=\frac{e^{-\theta_i}}{1+e^{-\theta_i}}$ is the
probability of connecting the node $i$ with any of the node of the opposite
layer L. The link probabilities are not uniform, but depend on the Lagrange multipliers of the nodes $i\in\Gamma$. The factors in the product in Eq.~\reff{eq:PBiPCM} express the 
probabilities of observing exactly the constrained node degrees: the probability of the degree $k_i$ of the node $i \in \Gamma$ is given by the probability of observing $k_i$ successes trials of a binomial distribution with probability $(p_\text{BiPCM})_i$. Maximizing the likelihood $\mathcal{L}$ returns the explicit expressions for the
link probabilities:
\begin{equation}
(p_\text{BiPCM})_i=\dfrac{k_i^*}{N_\Gamma}.
\end{equation}

\subsection{Bipartite Configuration Model}
In the monopartite configuration model, the degrees of all the nodes are constrained. Analogously, in the \emph{Bipartite Configuration Model} (BiCM,~\cite{Saracco2015a}) the degrees of the two layer degree sequences are constrained, such that $\langle
k_i\rangle=k_i^*,\,\forall i\in \Gamma$, and $\langle k_\alpha\rangle=k_\alpha^*,\,\forall
\alpha \in \text{L}$. If $\vec{\theta}$ and $\vec{\rho}$ are the corresponding Lagrange multipliers, the partition function reads~\cite{Saracco2015a}
\begin{equation}
\mathcal{Z}_\text{BiCM}(\vec{\theta}, \vec{\rho})=\prod_{i,\alpha}1+e^{-(\theta_i + \rho_\alpha)},
\end{equation}
following essentially the same strategy used in Eq.~\reff{eq:ZRG}. Again, the probability per graph factorizes in a product of probabilities per link:

\begin{equation}\label{eq:BiCMProbabilityDistribution}
\begin{split}
P(G_B|\vec{\theta}, \vec{\rho}) 
&=\prod_{i, \alpha} \dfrac{e^{-(\theta_i + \rho_\alpha)m_{i\alpha}}}{1+e^{-(\theta_i+\rho_\alpha)}}\\
&= \prod_{i, \alpha} (p_\text{BiCM})_{i\alpha}^{m_{i\alpha}}\big(1-(p_\text{BiCM})_{i\alpha}\big)^{1-m_{i\alpha}},
\end{split}
\end{equation}
where the probability per link reads
\begin{equation}
(p_\text{BiCM})_{i\alpha}=\frac{e^{-(\theta_i + \rho_\alpha)}}{1+e^{-(\theta_i+\rho_\alpha)}}, \quad i\in\Gamma, \alpha\in\text{L}
\label{eq:bicmProbability}
\end{equation}
Compared to the probability distributions of the BiRG and BiPCM, we can see that the BiCM distribution is more general and corresponds to the product of different Bernoulli events with link-specific success probabilities. Note that the distribution factorizes and link probabilities are independent. Maximizing the likelihood returns the equation system~\cite{Saracco2015a}
\begin{equation}
\left\{
\begin{array}{c}
\sum_\alpha\dfrac{e^{-(\theta_i+\rho_\alpha)}}{1+e^{-(\theta_i + \rho_\alpha)}}=k_i^*,\quad \forall i\in\Gamma,\\
\\
\sum_i\dfrac{e^{-(\theta_i+\rho_\alpha)}}{1+e^{-(\theta_i+\rho_\alpha)}}=k_\alpha^*,\quad \forall \alpha\in\text{L}.
\end{array}
\right.
\end{equation}
Solving this system allows us to evaluate the Lagrange multipliers and ultimately obtain the graph probabilities.

\subsection{Bipartite Weighted Configuration Model}
Constraining the node strengths as $\langle s_i \rangle = s^*_i, \forall i\in\Gamma$, and $\langle s_\alpha \rangle = s^*_\alpha, \forall \alpha\in$ L, give the \emph{Bipartite Weighted Configuration Model} (BiWCM,~\cite{diGangi2015}). Be $\vec{\theta}$ and $\vec{\rho}$ the corresponding Lagrange multipliers. As shown in~\cite{diGangi2015}, the partition function 
\begin{equation}
\mathcal{Z}_\text{BiCM}(\vec{\theta}, \vec{\rho}) = \prod_{i,\alpha} \frac{1}{	1 - e^{-(\theta_i +\rho_\alpha)}}.
\end{equation}
The graph probability yields
\begin{equation}
P(G_B|\vec{\theta}, \vec{\rho}) = \prod_{i, \alpha} \left(e^{-(\theta_i +\rho_\alpha)}\right)^{w_{i\alpha}} (1 - e^{-(\theta_i +\rho_\alpha)}).
\end{equation}
Similar to the BiCM, the Lagrange multipliers can be obtained by solving an equation system, which reads~\cite{diGangi2015}
\begin{equation}
\left\{
\begin{array}{c}
\sum_\alpha\dfrac{e^{-(\theta_i + \rho_\alpha)}}{1 - e^{-(\theta_i +\rho_\alpha)}} = s_i^*,\quad \forall i\in\Gamma,\\
\\
\sum_i\dfrac{e^{-(\theta_i +\rho_\alpha)}}{1 - e^{-(\theta_i + \rho_\alpha)}} = s_\alpha^*,\quad \forall \alpha\in\text{L}.\\
\end{array}
\right.
\end{equation}

\subsection{Bipartite Enhanced Configuration Model}
The \emph{Bipartite Enhanced Configuration Model} (BiECM, \cite{diGangi2015}) is a bipartite extension of the monopartite enhanced configuration model introduced in~\cite{Mastrandrea2014}. Both, degrees as well as strengths, are constrained.

Be $\theta_i$ and $\theta_\alpha$ the constraints associated to the degrees, and $\rho_i$ and $\rho_\alpha$ those associated to the strengths for the nodes $i\in\text{L}$ and $\alpha\in\Gamma$, respectively. Using the short-hand notation $\phi_i = e^{-\rho_i}, \xi_\alpha = e^{-\rho_\alpha}, \psi_i = e^{-\theta_i}$ and $\gamma_\alpha = e^{-\theta_\alpha}$, the partition function reads~\cite{diGangi2015}
\begin{equation}
\mathcal{Z}_{BiECM}(\vec{\theta}, \vec{\rho}) = \prod_{i, \alpha} \frac{1 - \phi_i \xi_\alpha ( 1 - \psi_i \gamma_\alpha)}{1 - \phi_i \xi_\alpha}.
\end{equation}
Consequently, the network probability is given by
\begin{equation}
P(G_B) = \prod_{i, \alpha} \frac{(1 - \phi_i \xi_\alpha) (\phi_i\xi_\alpha)^{w_{i\alpha}}(\psi_i\gamma_\alpha)^{\Theta(w_{i\alpha})}}{1 - \phi_i \xi_\alpha(1 - \psi_i\gamma_\alpha)}
\end{equation}
and factorizes in single link probabilities. The values of the Lagrange multipliers can be obtained through a nonlinear system of equations, as shown in the Appendix of~\cite{diGangi2015}.

\subsection{Maximum Entropy Capital Asset Pricing Model}

The elements of the weighted biadjacency matrix can be rescaled to yield the quantities of the \emph{Capital Asset Pricing Model} (CAPM,~\cite{Lintner1965,Mossin1966}). In the financial context, the vertex strengths are often described as the \emph{total asset size} of a bank (or \emph{market value} of their portfolio), $V_i = \sum_{\alpha} w_{i\alpha}$, and the \emph{market capitalization} of an asset, $C_\alpha = \sum_{i} w_{i\alpha}$~\cite{diGangi2015,Squartini2017}. In the CAPM, banks choose their portfolio weights proportional to their market value and the asset's capitalization:
\begin{equation}\label{eq:wCAPM}
w^{CAPM}_{i\alpha} = \frac{V_i C_\alpha}{w},
\end{equation}
where we have used $w = \sum_{i', \alpha'} w_{i'\alpha'}$
The probability distribution for the MECAPM yields~\cite{diGangi2015}
\begin{equation}
P(G_B) = \prod_{i, \alpha} \left[1 - \left(p_{CAPM}\right)_{i\alpha}\right]^{w_{i\alpha}} \left(p_{CAPM}\right)_{i\alpha},
\end{equation}
where the probability per link reads
\begin{equation}
\left(p_{CAPM}\right)_{i\alpha} = \frac{w^{CAPM}_{i\alpha}}{1 + w^{CAPM}_{i\alpha}}.
\end{equation}
Note that $P(G_B)$ is geometrically distributed for $w_{i\alpha} \in \mathbb{N}$~\cite{diGangi2015}. The link probabilities can be easily calculated using the identity in \eq{eq:wCAPM}.

\subsection{Enhanced Capital Asset Pricing Model}

The so-called \emph{Enhanced Capital Asset Pricing Model} (ECAPM,~\cite{Squartini2017}) reconstructs the link topology and subsequently the link weights. Their method makes only use of the strength sequence and is composed of two steps. 

Firstly, the topology of the network is reconstructed by using the BiCM under the assumption that the exponential Lagrange multipliers $x_i \equiv \text{e}^{-\theta_i}$ and $y_\alpha \equiv \text{e}^{-\theta_\alpha}$ are proportional to node-specific fitness values, represented by their strengths:
\begin{equation}
\begin{split}
x_i &\equiv \sqrt{z_\Gamma} s_i,\quad \forall i \in \Gamma\\
y_\alpha &\equiv \sqrt{z_L} s_\alpha,\quad \forall \alpha \in L
\end{split}
\label{eq:ecapmXY}
\end{equation}
Constraining the network density with the total number of links $\langle E\rangle \equiv E^*$, the parameter $z = \sqrt{z_\Gamma z_L}$ can be estimated using~\cite{Squartini2017}
\begin{equation}
\langle E \rangle = \sum_{i, \alpha} \frac{z V_i C_\alpha}{1 + z V_i C_\alpha},\quad \forall i \in \Gamma, \alpha \in L,
\end{equation}
Subsequently, the single link probabilities are simply given by the BiCM expression in Eq.~\reff{eq:bicmProbability}, substituting the Lagrange multipliers with the expressions~\reff{eq:ecapmXY}:
\begin{equation}
(p_{ECAPM})_{i\alpha} = \frac{z V_i C_\alpha}{1 + z V_i C_\alpha},\quad \forall i \in \Gamma, \alpha \in L,
\end{equation}
where $z$ absorbs the proportionality constants.

Secondly, the link weights are reconstructed using the CAPM model while taking the network topology into consideration. Instead of setting $w_{i\alpha} = V_i C_\alpha / w$, a correction factor is applied~\cite{Squartini2017}
\begin{equation}
\begin{split}
w_{i\alpha} 
&= m_{i\alpha} \frac{V_i C_\alpha}{w\ (p_{ECAPM})_{i\alpha}} \\
&= (V_i C_\alpha + z^{-1}) \frac{m_{i\alpha}}{w},
\end{split}
\end{equation}
where $m_{i\alpha}$ is 0 or 1,depending the link is present in the graph or not.

\bibliographystyle{spmpsci}
\bibliography{From_ecology_to_finance_and_back_recent_advancements}

\end{document}